\documentclass[sigconf]{acmart}

\usepackage{balance}

\def\BibTeX{{\rm B\kern-.05em{\sc i\kern-.025em b}\kern-.08emT\kern-.1667em\lower.7ex\hbox{E}\kern-.125emX}}
    
\newcommand{\cut}[1]{}
\newcommand{\chaowei}[1]{\textcolor{cyan}{[Chaowei: #1]}}

\copyrightyear{2019}
\acmYear{2019}
\setcopyright{acmcopyright}
\acmConference[CCS '19] {2019 ACM SIGSAC Conference on Computer and Communications Security}{November 11--15, 2019}{London, United Kingdom}
\acmBooktitle{2019 ACM SIGSAC Conference on Computer and Communications Security (CCS'19), November 11--15, 2019, London, United Kingdom}
\acmPrice{15.00}
\acmDOI{10.1145/3319535.3339815}
\acmISBN{978-1-4503-6747-9/19/11}

\usepackage{amsmath,amsfonts,bm}
\usepackage{adjustbox}
\usepackage[linesnumbered,ruled]{algorithm2e}

\usepackage{amsmath,amsfonts,bm}

\def\eqref#1{equation~\ref{#1}}

\def\1{\bm{1}}

\DeclareMathAlphabet{\mathsfit}{\encodingdefault}{\sfdefault}{m}{sl}
\SetMathAlphabet{\mathsfit}{bold}{\encodingdefault}{\sfdefault}{bx}{n}

\def\gL{{\mathcal{L}}}

\DeclareMathOperator*{\argmin}{arg\,min}

\newcommand{\adv}{\mathrm{adv}}
\newcommand{\advdelta}{T'}

\newcommand{\wx}{\mathbf{w_x}}
\newcommand{\wy}{\mathbf{w_y}}
\newcommand{\wz}{\mathbf{w_z}} 

\newcommand{\advdeltad}{t'}

\newcommand{\V}[1]{\mathbf{#1}}

\newcommand{\featuremap}{input feature matrix}

\newcommand{\sensor}{3D point cloud}
\newcommand{\advsensor}{spoofed 3D point cloud}
\newcommand{\pointcloud}{3D point cloud}
\newcommand{\spoof}{spoofed}
\newcommand{\advfeature}{spoofed input feature matrix}

\begin{document}
\fancyhead{}

\title{Adversarial Sensor Attack on LiDAR-based Perception in Autonomous Driving}

\author{Yulong Cao}
\affiliation{%
  University of Michigan
}
\email{yulongc@umich.edu}

\author{Chaowei Xiao}
\affiliation{%
  University of Michigan
}
\email{xiaocw@umich.edu}

\author{Benjamin Cyr}
\affiliation{%
  University of Michigan
}
\email{bencyr@umich.edu}

\author{Yimeng Zhou}
\affiliation{%
  University of Michigan
}
\email{yimzhou@umich.edu}

\author{Won Park}
\affiliation{%
  University of Michigan
}
\email{wonpark@umich.edu}

\author{Sara Rampazzi}
\affiliation{%
  University of Michigan
}
\email{srampazz@umich.edu}

\author{Qi Alfred Chen}
\affiliation{%
University of California, Irvine
}
\email{alfchen@uci.edu}

\author{Kevin Fu}
\affiliation{%
  University of Michigan
}
\email{kevinfu@umich.edu}

\author{Z. Morley Mao}
\affiliation{%
  University of Michigan
}
\email{zmao@umich.edu}

%

%
\begin{abstract}

In Autonomous Vehicles (AVs), one fundamental pillar is \emph{perception}, which leverages sensors like cameras and LiDARs (Light Detection and Ranging) to understand the driving environment. 
Due to its direct impact on road safety, multiple prior efforts have been made to study its the security of perception systems. In contrast to prior work that concentrates on camera-based perception, in this work we perform the first security study of LiDAR-based perception in AV settings, which is highly important but unexplored.
We consider LiDAR spoofing attacks as the threat model and set the attack goal as spoofing obstacles close to the front of a victim AV.
We find that blindly applying LiDAR spoofing is insufficient to achieve this goal due to the machine learning-based object detection process. Thus, we then explore the possibility of strategically controlling the spoofed attack to fool the machine learning model. We formulate this task as an optimization problem and design modeling methods for the input perturbation function and the objective function. We also identify the inherent limitations of directly solving the problem using optimization and design an algorithm that combines optimization and global sampling, which improves the attack success rates to around 75\%. As a case study to understand the attack impact at the AV driving decision level, we construct and evaluate two attack scenarios that may damage road safety and mobility. We also discuss defense directions at the AV system, sensor, and machine learning model levels.


\end{abstract}

%
%


%

%

%
\maketitle

\section{Introduction}
\label{sec:introduction}




Autonomous vehicles, or self-driving cars, are under rapid development, with some vehicles already found on public roads~\cite{waymo_public, lyft_public, baidu_apolong} In AV systems, one fundamental pillar is \textit{perception}, which leverages sensors like cameras and LiDARs (Light Detection and Ranging) to understand the surrounding driving environment. Since such function is directly related to safety-critical driving decisions such as collision avoidance, multiple prior research efforts have been made to study the security of camera-based perception in AV settings. For example, prior work has reported sensor-level attacks such as camera blinding~\citep{sensor-blackhat15}, physical-world camera attacks such as adding stickers to traffic signs~\citep{eykholt2018robust, eykholt2018physical}, and trojan attacks on the neural networks for AV camera input~\citep{liu2017trojaning}.




Despite the research efforts in camera-based perception, there is no thorough exploration into the security of LiDAR-based perception in AV settings.
LiDARs, which measure distances to surrounding obstacles using infrared lasers, can provide 360-degree viewing angles and generate 3-dimensional representations of the road environment instead of just 2-dimensional images for cameras. Thus, they are generally considered as more important sensors than cameras for AV driving safety~\citep{lidar_importance, lidar_intro} and are adopted by nearly all AV makers today~\citep{lidar_waymo, lidar_gm, lidar_volvo, apollo}. A few recent works demonstrated the feasibility of injecting spoofed points into the sensor input from the LiDAR~\citep{sensor-blackhat15, Shin2017IllusionAD}. Since such input also needs to be processed by an object detection step in the AV perception pipeline, it is largely unclear whether such spoofing can directly lead to semantically-impactful security consequences, e.g., adding spoofed road obstacles, in the LiDAR-based perception in AV systems.

In this work, we perform the first study to explore the security of LiDAR-based perception in AV settings. To perform the analysis, we target the LiDAR-based perception implementation in Baidu Apollo, an open-source AV system that has over 100 partners and has reached a mass production agreement with multiple partners such as Volvo and Ford~\citep{baidu_volvo_ford, baidu_apolong}. We consider a LiDAR spoofing attack, i.e., injecting spoofed LiDAR data points by shooting lasers, as our threat model since it has demonstrated feasibility in previous work~\citep{sensor-blackhat15, Shin2017IllusionAD}. With this threat model, we set the attack goal as adding spoofed obstacles in close distances to the front of a victim AV (or \textit{front-near} obstacles) in order to alter its driving decisions.



In our study, we first reproduce the LiDAR spoofing attack from the work done by \citet{Shin2017IllusionAD} and try to exploit Baidu Apollo's LiDAR-based perception pipeline, which leverages machine learning for object detection as with the majority of the state-of-the-art LiDAR-based AV perception techniques~\cite{kitti_3d}. We enumerate different spoofing patterns from the previous work, e.g., a spoofed wall, and different spoofing angles and shapes, but none of them succeed in generating a spoofed road obstacle after the machine learning step. We find that a potential reason is that the current spoofing technique can only cover a very narrow spoofing angle, i.e., 8$^\circ$ horizontally in our experiments, which is not enough to generate a point cloud of a road obstacle near the front of a vehicle. Thus, blindly applying existing spoofing techniques cannot easily succeed.

To achieve the attack goal with existing spoofing techniques, we explore the possibility of strategically controlling the spoofed points to fool the machine learning model in the object detection step. While it is known that machine learning output can be maliciously altered by carefully-crafted perturbations to the input~\cite{papernot2017practical, eykholt2018robust, carlini2016hidden, yuan2018commandersong, carlini2018audio}, no prior work studied LiDAR-based object detection models for AV systems. To approach this problem, we formulate the attack task as an optimization problem, which has been shown to be effective in previous machine learning security studies~\citep{carlini2017towards,cisse2017houdini,xie2017adversarial,xiao2018characterizing,cheng2018seq2sick,xiao2018generating}. Specific to our study, two functions need to be newly formulated: (1) an input perturbation function that models LiDAR spoofing capability in changing machine learning model input, and (2) an objective function that can reflect the attack goal. For the former, since previous work did not perform detailed measurements for the purpose of such modeling, we experimentally explore the capability of controlling the spoofed data points, e.g., the number of points and their positions. Next, we design a set of global spatial transformation functions to model these observed attack capabilities at the model input level. In this step, both the quantified attack capabilities and the modeling methodology are useful for future security studies of LiDAR-related machine learning models.


For the attack goal of adding front-near obstacles, designing a objective function is also non-trivial since the machine learning model output is post-processed in the perception module of Baidu Apollo before it is converted to a list of perceived obstacles. To address this, we study the post-processing logic, extract key strategies of transforming model output into perceived obstacles, and formulate it into the objective function.




With the optimization problem mathematically formulated, we start by directly solving it using optimization algorithms like previous studies~\citep{carlini2017towards}. However, we find that the average success rate of adding front-near obstacles is only 30\%. We find that this is actually caused by the nature of the problem, which makes it easy for any optimization algorithm to get trapped in local extrema. To solve this problem, we design an algorithm that combines global sampling and optimization, which is able to successfully increase the average success rates to around 75\%.



As a case study for understanding the impact of the discovered attack input at the AV driving decision level, we construct two attack scenarios: (1) \textit{emergency brake attack}, which may force a moving AV to suddenly brake and thus injure the passengers or cause rear-end collisions, and (2) \textit{AV freezing attack}, which may cause an AV waiting for the red light to be permanently ``frozen'' in the intersection and block traffic. Using real-world AV driving data traces released by the Baidu Apollo team, both attacks successfully trigger the attacker-desired driving decisions in Apollo's simulator.


Based on the insights from our security analysis, we propose defense solutions not only at AV system level, e.g., filtering out LiDAR data points from ground reflection, but also at sensor and machine learning model levels.


In summary, this work makes the following contributions:
\begin{itemize}
\setlength{\itemsep}{0pt}
\setlength{\parskip}{0pt}
    \item We perform the first security study of LiDAR-based perception for AV systems. We find that blindly applying existing LiDAR spoofing techniques cannot easily succeed in generating semantically-impactful security consequences after the machine learning-based object detection step. To achieve the attack goal with existing spoofing techniques, we then explore the possibility of strategically controlling the spoofed points to fool the machine learning model, and formulate the attack as an optimization problem.
    \item To perform analysis for the machine learning model used in LiDAR-based AV perception, we make two methodology-level contributions. First, we conduct experiments to analyze the LiDAR spoofing attack capability and design a global spatial transformation based method to model such capability in mathematical forms. Second, we identify inherent limitations of directly solving our problem using optimization, and design an algorithm that combines optimization and global sampling. This is able to increase the attack success rates to around 75\%.
    \item As a case study to understand the impact of the attacks at the AV driving decision level, we construct two potential attack scenarios: emergency brake attack, which may hurt the passengers or cause a rear-end collision, and AV freezing attack, which may block traffic. Using a simulation based evaluation on real-world AV driving data, both attacks successfully trigger the attacker-desired driving decisions. Based on the insights, we discuss defense directions at AV system, sensor, and machine learning model levels.
\end{itemize}
\section{Background}
\label{sec:background}

\subsection{LiDAR-based Perception in AV Systems}
\label{subsec:LiDAR_object_detection}
\begin{figure*}[t]
  \centering
    \includegraphics[width=0.95\textwidth]{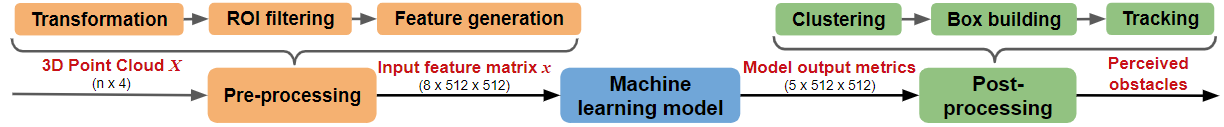}
    \vspace{-0.1in}
    \caption{Overview of the data processing pipeline for LiDAR-based perception in Baidu Apollo.}
    \vspace{-0.1in}
    \label{fig:LiDAR_object_detection}
\end{figure*}

AVs rely on various sensors to perform real-time positioning (also called localization) and environment perception (or simply perception). LiDAR, camera, radar, and GPS/IMU are major sensors used by various autonomous driving systems. The data collected from those sensors are transformed and processed before it becomes useful information for AV systems. 
Fig.~\ref{fig:LiDAR_object_detection} shows the data processing pipeline of LiDAR sensor data in the perception module of Baidu Apollo~\cite{apollo}. As shown, it involves three main steps as follows:


\textbf{Step 1: Pre processing}. The raw LiDAR sensor input is called \textit{\pointcloud} and we denote it as $X$. The dimension of $X$ is $n \times 4$, where $n$ denotes the number of data points and each data point is a 4-dimension vector with the 3D coordinates, $\wx$, $\wy$, and $\wz$, and the intensity of the point. In the pre-processing step, $X$ is first transformed into an absolute coordinate system.
Next, the \emph{Region of Interest (ROI)} filter removes unrelated portions of the \pointcloud{} data, e.g., those that are outside of the road, based on HDMap information. 
Next, 
a \emph{feature generation} process generates a feature matrix $x$ $(8 \times 512 \times 512)$, which is the input to the subsequent machine learning model. In this process, the ROI-filtered \pointcloud{} within the range (60 meters by default) is mapped to $512 \times 512$ cells according to the $\wx$ and $\wy$ coordinates. In each cell, the assigned points are used to generate $8$ features as listed in Table~\ref{table:cnn_input}.

\textbf{Step 2: DNN-based object detection.} A Deep Neural Network (DNN) then takes the feature matrix $x$ as input and produces a set of output metrics for each cell, e.g., the probability of the cell being a part of an obstacle. These output metrics are listed in Table \ref{table:cnn_output}.  

\textbf{Step 3: Post processing.} 
The clustering process only considers cells with \emph{objectness} values (one of the output metrics listed in Table \ref{table:cnn_output}) greater than a given threshold ($0.5$ by default). Then, the process constructs candidate object clusters by building a connected graph using the cells' output metrics. Candidate object clusters are then filtered by selecting clusters with average \emph{positiveness} values (another output metric) greater than a given threshold ($0.1$ by default). The \emph{box builder} then reconstructs the bounding box including height, width, length of an obstacle candidate from the \pointcloud{} assigned to it. Finally, the \emph{tracker} integrates consecutive frames of processed results to generate tracked obstacles, augmented with additional information such as speed, acceleration, and turning rates, as the output of the LiDAR-based perception.


With the information of perceived obstacles such as their positions, shapes, and obstacle types, the Apollo system then uses such information to make driving decisions. The perception output is further processed by the \emph{prediction} module which predicts the future trajectories of perceived obstacles, and then the \emph{planning} module which plans the future driving routes and makes decisions such as stopping, lane changing, yielding, etc.

\begin{table}[h]\footnotesize
\begin{tabular}{| l | p{6.0cm} |}
\hline
\textbf{Feature} & \textbf{Description} \\
\hline
\textbf{Max height} & Maximum height of points in the cell. \\
\hline
\textbf{Max intensity} & Intensity of the highest point in the cell. \\
\hline
\textbf{Mean height} & Mean height of points in the cell. \\
\hline
\textbf{Mean intensity} & Mean intensity of points in the cell. \\
\hline
\textbf{Count} & Number of points in the cell. \\
\hline
\textbf{Direction} & Angle of the cell's center with respect to the origin. \\
\hline
\textbf{Distance} & Distance between the cell's center and the origin. \\
\hline
\textbf{Non-empty} & Binary value indicating whether the cell is empty or occupied. \\
\hline
\end{tabular}
\\
  \caption{DNN model input features.}
  \label{table:cnn_input}
\end{table}

\begin{table}[h]\footnotesize
\begin{tabular}{| l | p{5.7cm} |}
\hline
\textbf{Metrics} & \textbf{Description} \\
\hline
\textbf{Center offset} & Offset to the predicted center of the cluster the cell belongs to. \\
\hline
\textbf{Objectness} & The probability of a cell belonging to an obstacle. \\
\hline
\textbf{Positiveness} & The confidence score of the detection. \\
\hline
\textbf{Object height} & The predicted object height. \\
\hline
\textbf{Class probability} & The probability of the cell being a part of a vehicle, pedestrian, etc. \\
\hline
\end{tabular}
\\
  \caption{DNN model output metrics.}
  \label{table:cnn_output}
\end{table}

\subsection{LiDAR Sensor and Spoofing Attacks} ~\label{sec:lidar-spoofing}

To understand the principles underlying our security analysis methodology, it is necessary to understand how the LiDAR sensor generates a point cloud and how it is possible to alter it in a controlled way using spoofing attacks.

\textbf{LiDAR sensor.} A LiDAR sensor functions by firing laser pulses and capturing their reflections using photodiodes. Because the speed of light is constant, the time it takes for the echo pulses to reach the receiving photodiode provides an accurate measurement of the distance between a LiDAR and a potential obstacle. By firing the laser pulses at many vertical and horizontal angles, a LiDAR generates a point cloud used by the AV systems to detect objects.

\textbf{LiDAR spoofing attack.} Sensor spoofing attacks use the same physical channels as the targeted sensor to manipulate the sensor readings. This strategy makes it very difficult for the sensor system to recognize such attack, since the attack doesn't require any physical contact or tampering with the sensor, and it doesn't interfere with the processing and transmission of the digital sensor measurement. These types of attacks could trick the victim sensor to provide seemingly legitimate, but actually erroneous, data.

LiDAR has been shown to be vulnerable to laser spoofing attacks in prior work. Petit et al. demonstrated that a LiDAR spoofing attack can be performed by replaying the LiDAR laser pulses from a different position to create fake points further than the location of the spoofer~\cite{sensor-blackhat15}. Shin et al. showed that it is possible to generate a fake point cloud at different distances, even closer than the spoofer location~\cite{Shin2017IllusionAD}. In this paper, we build upon these prior works to study the effect of this attack vector on the security of AV perception.




\subsection{Adversarial Machine Learning}
\textbf{Neural networks.}
A neural network is a function consisting of connected units called (artificial) neurons that work together to represent a differentiable function that outputs a distribution.
A given neural network (e.g., classification) can be defined by its model architecture and parameters $\phi$. An optimizer such as Adam~\cite{kingma2014adam} is used to update the parameters $\phi$ with respect to the objective function $\gL$. 

\textbf{Adversarial examples.}
Given a machine learning model $M$, input $x$ and its corresponding label $y$, an adversarial attacker aims to generate adversarial examples $x'$ so that $M(x')\neq y$ (untargeted attack) or $M(x') = y'$, where $y'$ is a target label (targeted attack). \citet{carlini2017towards} proposed to generate an adversarial perturbation for a targeted attack by optimizing an objective function as follows:
\begin{equation}
\label{eq:aml_opt}
\min ||x - x'||_p \quad \text{s.t.} \qquad M(x') = y' \nonumber  \quad \text{and}  \quad  x' \in X \nonumber,
\end{equation}
where $M(x')=y'$ is the target adversarial goal and $x'\in X$ denote that the adversarial examples should be in a valid set. Further, optimization-based algorithms have been leveraged to generate adversarial examples on various kinds of machine learning tasks successfully, such as segmentation~\cite{xie2017adversarial,cisse2017houdini}, human pose estimation~\cite{cisse2017houdini}, object detection~\cite{xie2017adversarial}, Visual Question Answer system~\cite{xu2017can}, image caption translation~\cite{cheng2018seq2sick}, etc. In this paper, we also leverage an optimization-based method to generate adversarial examples to fool LiDAR-based AV perception.  



\vspace{-\topsep}

\section{Attack Goal and Threat Model}
\label{sec:thread_model}
\textbf{Attack goal.} To cause semantically-impactful security consequences in AV settings, we set the attack goal as fooling the LiDAR-based perception into perceiving fake obstacles in front of a victim AV in order to maliciously alter its driving decisions. More specifically, in this work, we target \textit{front-near} fake obstacles, i.e., those that are in close distances to the front of a victim AV, since they have the highest potential to trigger immediate erroneous AV driving decisions. In this work, we define front-near obstacles as those that are around 5 meters in front of a victim AV.



\textbf{Threat model.} To achieve the attack goal above, we consider LiDAR spoofing attacks as our threat model, which is a demonstrated practical attack vector for LiDAR sensors~\citep{sensor-blackhat15,Shin2017IllusionAD} as described in~\S\ref{sec:lidar-spoofing}. In AV settings, there are several possible scenarios to perform such attack. First, the attacker can place an attacking device at the roadside to shoot malicious laser pulses to AVs passing by. Second, the attacker can drive an attack vehicle in close proximity to the victim AV, e.g., in the same lane or adjacent lanes. To perform the attack, the attack vehicle is equipped with an attacking device that shoots laser pulses to the victim AV's LiDAR. To perform laser aiming in these scenarios, the attacker can use techniques such as camera-based object detection and tracking. In AV settings, these attacks are stealthy since the laser pulses are invisible and laser shooting devices are relatively small in size. 

As a first security analysis, we assume that the attacker has white-box access to the machine learning model and the perception system. We consider this threat model reasonable since the attacker could obtain white-box access by additional engineering efforts to reverse engineering the software.

\begin{figure*}[t]
  \centering
    \includegraphics[width=0.9\textwidth]{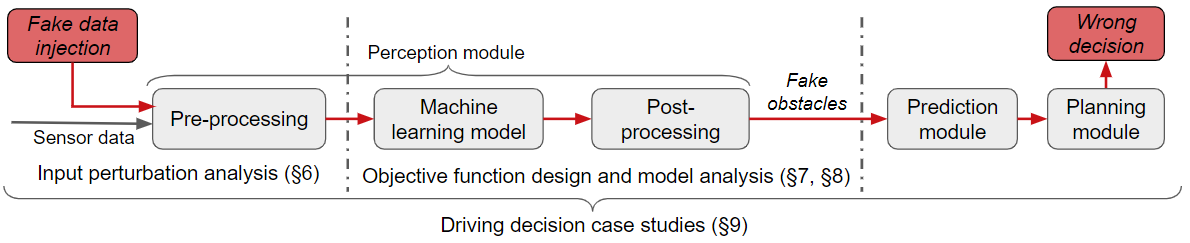}
    \caption{Overview of the Adv-LiDAR methodology.
    }
    \label{fig:method_overview}
\end{figure*}

\section{Limitation of Blind sensor spoofing}
\label{sec:spoofing-attack}



To understand the security of LiDAR-based perception under LiDAR spoofing attacks, we first reproduce the state-of-the-art LiDAR spoofing attack by Shin et al.~\cite{Shin2017IllusionAD}, and explore the effectiveness of directly applying it to attack the LiDAR-based perception pipeline in Baidu Apollo~\cite{apollo}, an open-source AV system that has over 100 partners and has reached mass production agreements with multiple partners such as Volvo, Ford, and King Long~\citep{baidu_volvo_ford, baidu_apolong}.

\begin{figure}[t]
  \centering
    \includegraphics[width=0.4\textwidth]{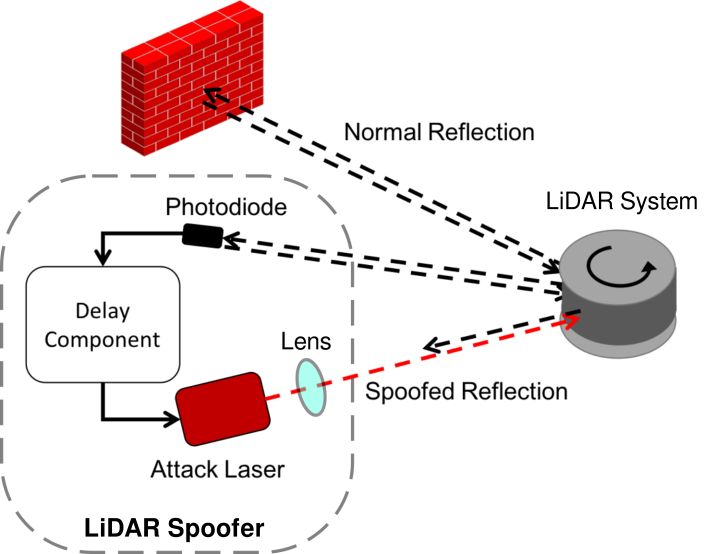}
    \caption{Illustration of LiDAR spoofing attack. The photodiode receives the laser pulses from the LiDAR and activate the delay component that triggers the attacker laser to simulate real echo pulses.}
    \label{fig:sensor_attack}
\end{figure}

\textbf{Spoofing attack description.}
The attack by Shin et al.~\cite{Shin2017IllusionAD} consists of three components: a photodiode, a delay component, and an infrared laser, which are shown in Fig.~\ref{fig:sensor_attack}. The photodiode is used to synchronize with the victim LiDAR. The photodiode triggers the delay component whenever it captures laser pulses fired from the victim LiDAR. Then the delay component triggers the attack laser after a certain amount of time to attack the following firing cycles of the victim LiDAR. Since the firing sequence of laser pulses is consistent, an adversary can choose which fake points will appear in the point cloud by crafting a pulse waveform to trigger the attack laser.

\begin{figure}
  \centering
    \includegraphics[width=0.4\textwidth]{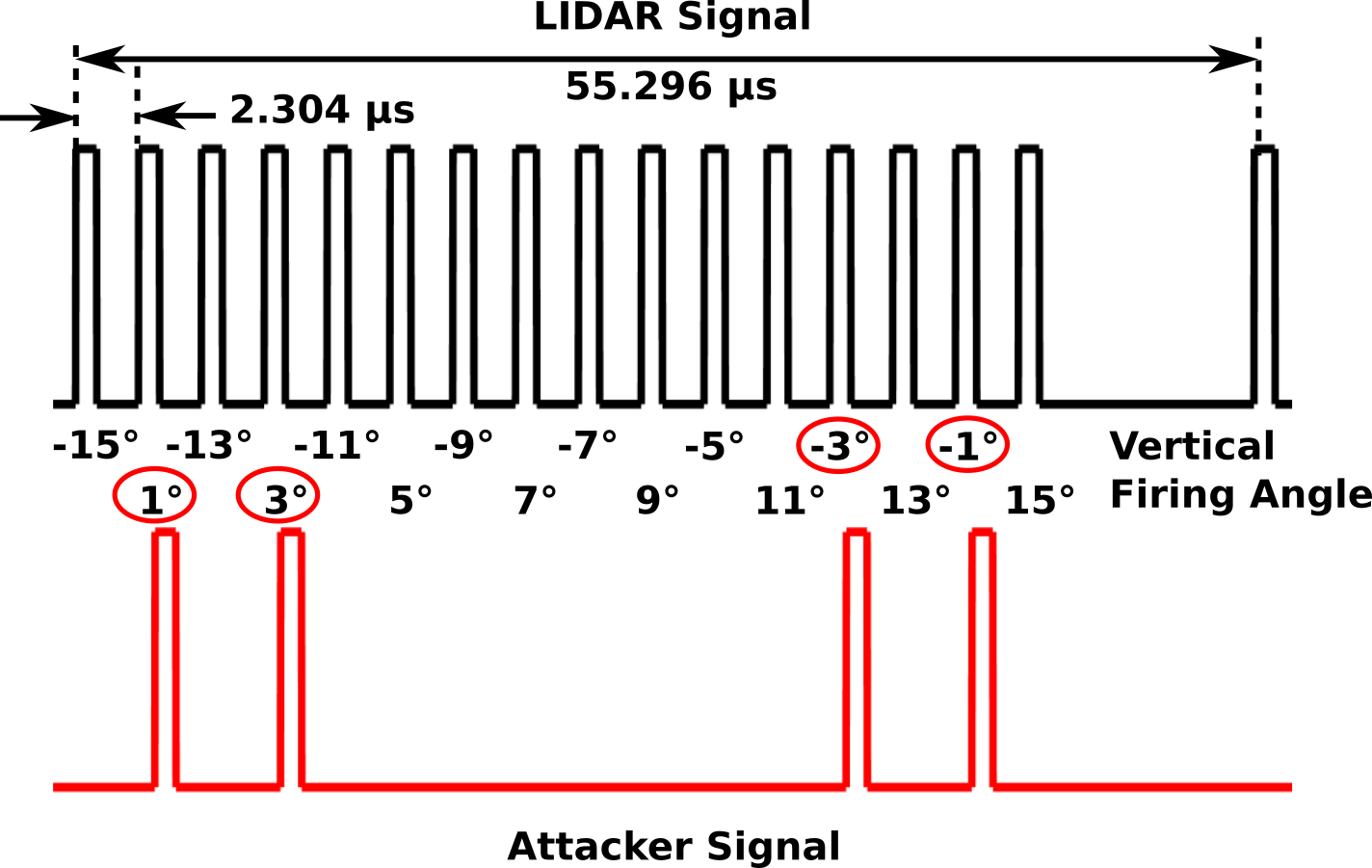}
    \caption{The consistent firing sequence of the LiDAR allows an attacker to choose the angles and distances from which spoofed points appear. For example, applying the attacker signal, fake dots will appear at 1$^\circ$, 3$^\circ$, -3$^\circ$, and -1$^\circ$ angles (0$^\circ$ is the center of the LiDAR)}
    \label{fig:firing_sequence}
\end{figure}

\textbf{Experimental setup.}
We perform spoofing attack experiments on a VLP-16 PUCK LiDAR System from Velodyne~\cite{vpl}. The VLP-16 uses a vertical array of 16 separate laser diodes to fire laser pulses at different angles. It has a 30 degree vertical angle range from -15 $^\circ$ to +15 $^\circ$, with 2 $^\circ$ of angular resolution. The VLP-16 rotates horizontally around a center axis to send pulses in a 360 $^\circ$ horizontal range, with a varying azimuth resolution between 0.1 $^\circ$ and 0.4  $^\circ$. The laser firing sequence follows the pattern shown in Figure~\ref{fig:firing_sequence}. The VLP-16 fires 16 laser pulses in a cycle every 55.296 $\mu$s, with a period of 2.304 $\mu$s. The receiving time window is about 667 ns. We chose this sensor because it is compatible with Baidu Apollo and uses the same design principle as the more advanced HDL-64E LiDARs used in many AVs. The similar design indicates that the same laser attacks that affect the VLP-16 can be extended to high-resolution LiDARs like the HDL-64E. 

We use the OSRAM SFH 213 FA as our photodiode, with a comparator circuit similar to the one used by Shin et al. We use a Tektronix AFG3251 function generator as the delay component with the photodiode circuit as an external trigger. In turn, the function generator provides the trigger to the laser driver module PCO-7114 that drives the attack laser diode OSRAM SPL PL90. With the PCO-7114 laser driver, we were able to fire the laser pulses at the same pulse rate of the VLP-16, 2.304 $\mu$s, compared to 100 $\mu$s of the previous work. An optical lens with a diameter of 30mm and a focal length of 100 mm was used to focus the beam, making it more effective for ranges farther than 5 meters. We generate the custom pulse waveform using the Tektronix software ArbExpress \cite{arbexpress} to create different shapes and the Velodyne software VeloView \cite{veloview} to analyze and extract the point clouds.

\textbf{Experiment results.}
The prior work of Shin et al. is able to spoof a maximum of 10 fake dots in a single horizontal line. With our setup improvements (a faster firing rate and a lens to focus the beam), fake points can be generated at all of the 16 vertical viewing angles and an 8 $^\circ$ horizontal angle at greater than 10 meters away. In total, around 100 dots can be spoofed by covering these horizontal and vertical angles (illustrated in Fig.~\ref{fig:LiDAR_sensor_attack} in Appendix). These spoofed dots can also be shaped by modifying the custom pulse waveform used to fire the attack laser. Notice that even though around 100 dots can be spoofed, they are not all spoofed stably. The attacker is able to spoof points at different angles because the spoofed laser pulses hit a certain area on the victim LiDAR due to the optical lens focusing. The closer to the center of the area, the stronger and stabler laser pulses are received by the victim LiDAR. We find that among 60 points at the center 8-10 vertical lines can be stably spoofed with high intensity.

\subsection{Blind LiDAR Spoofing Experiments}

\begin{figure}
  \centering
    \includegraphics[width=0.47\textwidth]{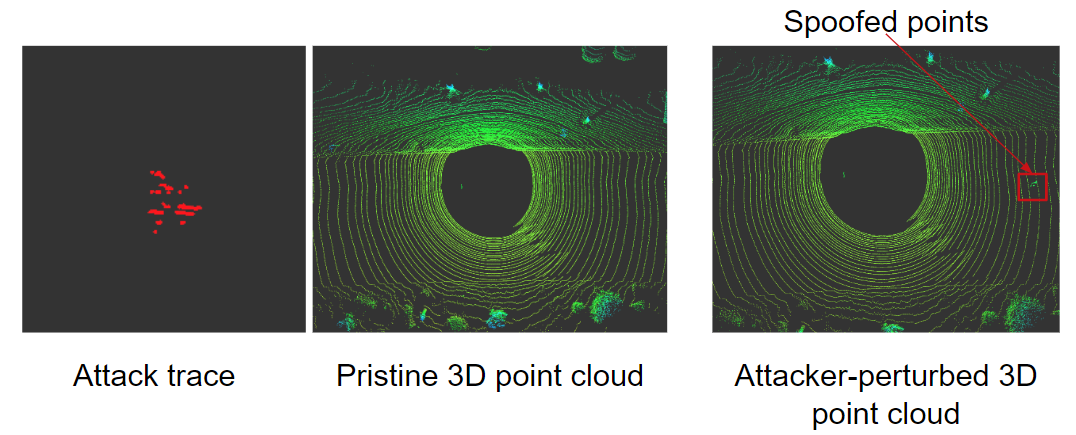}
    \caption{Generating the attacker-perturbed 3D point cloud by synthesizing the pristine 3D point cloud with the attack trace to spoof a front-near obstacle 5 meters away from the victim AV.}
    \label{fig:sensor-baseline}
\end{figure}

After reproducing the LiDAR spoofing attack, we then explore whether blindly applying such attack can directly generate spoofed obstacles in the LiDAR-based perception in Baidu Apollo. Since our LiDAR spoofing experiments are performed in indoor environments, we synthesize the on-road attack effect by adding spoofed LiDAR points to the original \pointcloud{} collected by Baidu Apollo team on local roads in Sunnyvale, CA. The synthesizing process is illustrated in Fig.~\ref{fig:sensor-baseline}. After this process, we run Apollo's perception module with the attacker-perturbed \pointcloud{} as input to obtain the object detection output. In this analysis, we explore three blind attack experiments as follows:


\textbf{Experiment 1: Directly apply original spoofing attack traces.}
In this experiment, we directly replay spoofing attack traces to attack LiDAR-based perception in Apollo. More specifically, we experiment with attack traces obtained from two sources: (1) the original spoofing attack traces from Shin et al.~\cite{Shin2017IllusionAD}, and (2) the attack traces generated from the spoofing attack reproduced by us, which can inject more dots after our setup improvements. However, we are not able to observe a spoofed obstacle for any of these traces at the output of the LiDAR-based perception pipeline.



\textbf{Experiment 2: Apply spoofing attack traces at different angles.}
To understand whether successfully spoofing an obstacle depends on the angle of the spoofed points, in this experiment we inject spoofed points at different locations. More specifically, we uniformly sample 100 different angles out of 360 degrees around the victim AV, and inject the spoofing attack traces reproduced by us. However, we are not able to observe spoofed obstacles for any of these angles.


\textbf{Experiment 3: Apply spoofing attack traces with different shapes.}
To understand whether successfully spoofing an obstacle depends on the pattern of the spoofed points, in this experiment we inject points with different spoofing patterns. More specifically, we generate random patterns of spoofed points by randomly setting distances for each point at different angles. We generate 160 points covering 16 vertical lines, 10 points for each line with continuous horizontal angles. To trigger immediate control decision changes in an AV, the spoofed obstacle needs to be close to the victim AV. Thus, we set the generated distances of the spoofed point to be within 4 to 6 meters to the victim AV. We generate 100 different spoofed patterns in total, but we are not able to observe spoofed obstacles for any of these patterns.


\textbf{Summary.} In these experiments, we try various blind spoofing attack strategies directly derived from the state-of-the-art LiDAR spoofing attack, but none of them succeed in generating spoofed obstacles in the LiDAR-based perception pipeline in Baidu Apollo. There are two potential reasons. First, as described earlier, the current attack methodology can only cover a very narrow spoofing angle, i.e., 8 $^\circ$ of horizontal angle even after our setup improvements. Second, the coverage of vertical angles is limited by the frequency of spoofing laser pulses. Thus, when attacking a LiDAR with more vertical angles, e.g., a 64-line LiDAR, since a 64-line LiDAR takes similar time as a 16-line LiDAR in scanning vertical angles, the attacker cannot spoof more vertical angles than those for a 16-line LiDAR. Thus, the current methodology limits the number of spoofed points, making it hard to generate enough points to mimic an important road obstacle.

\begin{figure}
  \centering
    \includegraphics[width=0.4\textwidth]{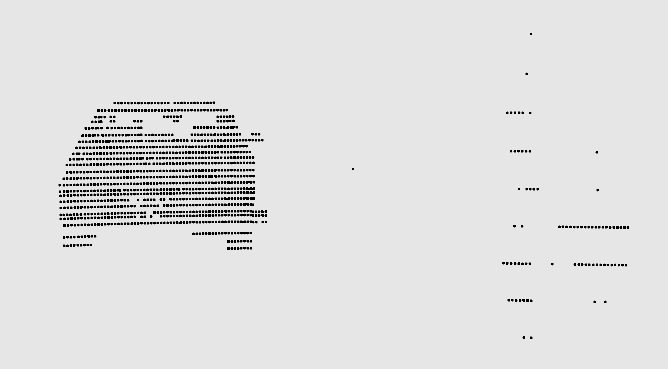}
    \caption{The point cloud from a real vehicle reflection (left) and from the spoofing attack (right) in a 64-line HDL-64E LiDAR. The vehicle is around 7 meters in front of the AV. }
    \label{fig:side-compare}
\end{figure}

To illustrate that, as shown in Fig.~\ref{fig:side-compare}, the point cloud for a real vehicle has a much wider angle and many more points than the attack traces reproduced by us. Thus, blindly applying the spoofing attack cannot easily fool the machine learning based object detection process in the LiDAR-based perception pipeline. In the next section, we explore the possibility of further exploiting machine learning model vulnerabilities to achieve our attack goal.



\section{Improved Methodology: Adv-LiDAR}
\label{sec:overview}

As discussed in~\S\ref{sec:spoofing-attack}, without considering the machine learning model used in LiDAR-based perception, blindly applying existing LiDAR spoofing attacks can hardly achieve the attack goal of generating front-near obstacles. Since it is known that machine learning output can be maliciously altered by carefully-crafted perturbations to the input~\cite{papernot2017practical, eykholt2018robust, carlini2016hidden, yuan2018commandersong, carlini2018audio}, we are then motivated to explore the possibility of strategically controlling the spoofed points to fool the machine learning model in LiDAR-based perception. In this section, we first describe the technical challenges after involving adversarial machine learning analysis in this research problem, and then present our solution methodology overview, called \textit{Adv-LiDAR}.

\subsection{Technical Challenges}
\label{sec:challenges}

Even though previous studies have shown promising results in attacking machine learning models, none of them studied LiDAR-based object detection models, and their approaches have limited applicability to our analysis goal due to three challenges:
    
First, attackers have limited capability of perturbing machine learning model inputs in our problem. Other than perturbing pixels on an image, perturbing machine learning inputs under AV settings requires perturbing 3D point cloud raw data by sensor attack and bypassing the associated pre-processing process. Therefore, such perturbation capability needs to be quantified and modeled.

Second, optimization-based methods for generating adversarial examples in previous studies may not be directly suitable for our analysis problem due to the limited model input perturbation capability. As shown in~\S\ref{sec:aml_analysis}, we find that optimization-based methods are inherently limited due to the nature of our problem, and can only achieve very low success rate in generating front-near obstacles. 

 Third, in our problem, successfully changing the machine learning model output does not directly lead to successes in achieving our attack goal in AV settings. As detailed later in~\S\ref{sec:aml_analysis}, in AV systems such as Baidu Apollo, machine learning model output is post-processed before it is converted to a list of perceived obstacles. Thus, an objective function that can effectively reflect our attack goal needs to be newly designed.
    

\subsection{Adv-LiDAR Methodology Overview}

\begin{table*}[h]\footnotesize
\centering
\begin{adjustbox}{max width=\linewidth}
\begin{tabular}{| l | l | l | l |}
\hline
\textbf{Notation} & \textbf{Description} &\textbf{Notation} & \textbf{Description} \\
\hline
$X$ & \pointcloud & $x$ & Input feature matrix\\
$X'$ & Adversarial \pointcloud & $x'$ & Adversarial \featuremap\\
$T$ &  Spoofed \pointcloud & $t$ & Spoofed \featuremap\\
$T'$ & Adversarial \spoof{} \pointcloud & $t'$ & Adversarial \spoof{} \featuremap\\
$(\wx,\wy,\wx)$ &3D Cartesian coordinate& $L_{\theta}, L_{\tau}$ & Upper bound of $\theta, \tau$ during sampling\\
$(u,v)$ & Coordinate of $t$& $(u',v')$ & Coordinate of $t'$\\
$M$ & Machine learning model & $I_{\cdot}$ & Model outputs \\
$N(u,v)$ & 4-pixel neighbor at the location $(u,v)$  & $S(\cdot)$ & Height Scaling function \\
$\mathcal{A}$ &  Spoofing attack capability & $\Phi(\cdot)$ & Mapping function (3D$\rightarrow$ 2D)\\
$Q(M, \cdot)$ & Extraction function& $\oplus(\cdot)$ & Merge function \\
$\mathcal{M}(\cdot)$ & Gaussian mask& $(px, py)$ & Center points of the Gaussian mask\\

$f(\cdot)$ & Objective function& $\gL_{\adv}(\cdot)$ & Adversarial loss \\ 
$H(\theta, \tau, \epsilon)$ & 2D Homography Matrix ($\theta$ : rotation, $\epsilon$ : scaling ; $\tau$ : translation )   & $S_{h}$  & Height scaling ratio\\
$\mathcal{S_\mathrm{T}}$ & Set of spoofed \pointcloud &  $\mathcal{S_\mathrm{t}}$ & Set of spoofed \featuremap \\
$G_T(T,\cdot)$  & Global spatial transformation function for \sensor     & $G_t(t,\cdot)$  & Global spatial transformation function for \featuremap \\
\hline

\end{tabular}
\end{adjustbox}
  \caption{Notations adopted in this work.}
  \label{table:notation}
\end{table*}

In this section, we provide an overview of our solution methodology, which we call Adv-LiDAR, that addresses the three challenges above. At a high level, to identify adversarial examples for the machine learning model $M$, we adopt an optimization-based approach, which has shown both high efficiency and effectiveness by previous studies for machine learning models across different domains~\cite{carlini2017towards,cisse2017houdini,xiao2018spatially,xiao2018generating}. To help explain the formulation of the optimization problem, we summarize the notations in Table~\ref{table:notation}. Specifically, the problem is formulated as follows:
\begin{equation}
\label{eq:aml_goal_opt}
\begin{split}
    & \min \qquad \gL_{\adv} (x \oplus \advdeltad;M) \qquad
    \\ & \text{s.t.} \qquad \advdeltad \in  \{ \Phi(\advdelta) | \advdelta \in \mathcal{A} \} \And x = \Phi(X)
\end{split}
\end{equation}
where $X$ is the pristine \sensor{} and $x$ represents the corresponding 2D \featuremap.  $\Phi(\cdot)$ is the pre-processing function that maps $X$ into $x$ (\S\ref{subsec:LiDAR_object_detection}). $\advdelta$ and $\advdeltad$ are the corresponding adversarial \advsensor{} and adversarial \advfeature. 
$\mathcal{A}$ is a set of \advsensor{} generated from LiDAR spoofing attacks. $\gL_{\adv}(\cdot; M)$ is the adversarial loss designed to achieve the adversarial goal given the machine learning model $M$. The constraints are used to guarantee that the generated adversarial examples $\advdeltad$ satisfy the spoofing attack capability.






Figure~\ref{fig:method_overview} overviews the analysis tasks needed to solve the optimization problem. First, we need to conduct an input perturbation analysis that formulates the spoofing attack capabilities $\mathcal{A}$ and merging function $\oplus$. Second, we need to perform a model analysis to design an objective function to generate adversarial examples. Third, as a case study to understand the impact of the attacks at the AV driving decision level, we further perform a driving decision analysis using the identified adversarial examples. More details about these tasks are as follows:

\textbf{Input perturbation analysis.} Formulating $\mathcal{A}$ and $\oplus$ is non-trivial. First, previous work regarding LiDAR spoofing attacks neither provided detailed measurements on the attacker's capability in perturbing \sensor{} nor expressed it in a closed form expression. 
Second, point cloud data is pre-processed by several steps as shown in Section~\ref{subsec:LiDAR_object_detection} before turning into machine learning input, which means the merging function $\oplus$ cannot be directly expressed.
To address these two challenges, as will be detailed later in~\S\ref{sec:input_perturb}, we first conduct spoofing attacks on LiDAR to collect a set of possible \advsensor{}. Using such \advsensor{}, we model the spoofing attack capability $\mathcal{A}$. We further analyze the pre-processing program to obtain the additional constraints to the machine learning input perturbation, or the \advfeature{}. Based on this analysis, we formulate the \advfeature{} into a differentiable function using global spatial transformations, which is required for the model analysis.

\textbf{Objective function design and model analysis.} As introduced earlier in~\S\ref{sec:challenges}, in LiDAR-based perception in AV systems, the machine learning model output is post-processed (\S~\ref{subsec:LiDAR_object_detection}) before turning into a list of perceived obstacles. To find an effective objective function, we study the post-processing steps to extract key strategies of transforming model output into perceived obstacles, and formulate it into an objective function that reflects the attack goal. 
In addition, we find that our optimization problem cannot be effectively solved by directly using existing optimization-based methods. We analyze the loss surface, and find that this inefficiency is caused by the problem nature. To address this challenge, we improve the methodology by combining global sampling with optimization. Details about the analysis methodology and results are in~\S\ref{sec:aml_analysis} and~\S~\ref{sec:eval}.

\textbf{Driving decision case study.} With the results from previous analysis steps, we can generate adversarial \pointcloud{} that can inject spoofed obstacles at the LiDAR-based perception level. To understand their impact at the AV driving decision level, we construct and evaluate two attack scenarios as case studies. The evaluation methodology and results are detailed later in~\S\ref{sec:e2e}.
\section{Input Perturbation Analysis}
To generate adversarial examples by solving the above optimization problem in Equation~\ref{eq:aml_opt}, we need to formulate merging function $\oplus$ and \featuremap{} spoofing capability $\Phi(\mathcal{A})$ as a closed form. In this section, we first analyze the spoofing attack capability ($\mathcal{A}$), 
and then use it to formulate $\Phi(\mathcal{A})$.
\label{sec:input_perturb}

\subsection{Spoofing Attack Capability}
\label{subsec:sensor_data_perturb}

Based on the attack reproduction experiments in ~\S\ref{sec:spoofing-attack}, the observed attack capability ($\mathcal{A}$) can be described from two aspects:

\textbf{Number of spoofed points}. As described in~\S\ref{sec:spoofing-attack}, even though it is possible to spoof around 100 points after our setup improvement, we find that around 60 points can be reliably spoofed in our experiments. Thus, we consider 60 as the highest number of reliable spoofed points. Noticed that, the maximum number of spoofed points could be increased if the attacker uses more advanced attack equipment. Here, we choose a set of devices that are more accessible (detailed in ~\S\ref{sec:spoofing-attack}) and end up with the ability to reliably spoof around 60 points. In addition, considering that an attacker may use a slower laser or cruder focusing optics, such as in the setup by Shin et al.~\cite{Shin2017IllusionAD}, we also consider 20 and 40 spoofed points in our analysis. 

\begin{figure}
  \centering
    \includegraphics[width=0.45\textwidth]{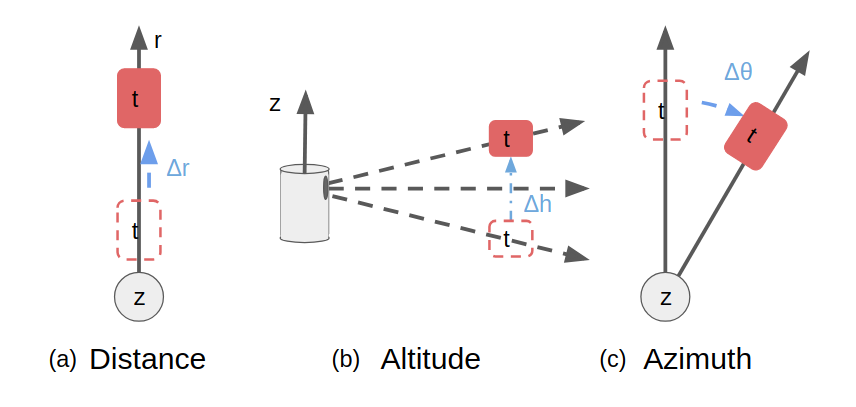}
    \caption{Attack capability in perturbing 3D Point Cloud $T$ }
    \label{fig:perturb_capabilities}
\end{figure}

\textbf{Location of spoofed points}. Given the number of spoofed points, the observed attack capability in placing these points are described and modeled as follows:
\begin{enumerate}
    \item Modify the {\em distance} of the spoofed point from the LiDAR by changing the delay of the attack laser signal pulses in small intervals (nanosecond scale). From the perspective of \advsensor{} $T$, this can be modeled as moving the position of the spoofed points nearer or further on the axis $r$ that connects the spoofed points and the LiDAR sensor by distance $\Delta r$ (Fig.~\ref{fig:perturb_capabilities}~(a)).
    \item Modify the {\em altitude} of a spoofed point within the vertical range of the LiDAR by changing the delay in intervals of 2.304 $\mu{}s$. From the perspective of \advsensor{} $T$, this can be modeled as moving the position of the spoofed points from vertical line to vertical line to change the height of it by height $\Delta h$ (Fig.~\ref{fig:perturb_capabilities}~(b)).
    \item Modify the {\em azimuth} of a spoofed point within a horizontal viewing angle of 8$^\circ$ by changing the delay in intervals of 55.296 $\mu{}s$. By moving the LiDAR spoofer to different locations around the LiDAR, it is possible to spoof at any horizontal angle. From the perspective of \advsensor{} $T$, this can be modeled as rotating the spoofed points with the LiDAR sensor as the pivot point on the horizontal plane by angle $\Delta \theta$ (Fig.~\ref{fig:perturb_capabilities}~(c)).
\end{enumerate}

Therefore, 
we model the attack capability $\mathcal{A}$ by applying these three modifications to the given \advsensor{} $T$. 
Here the \advsensor{} is collected by reproducing the sensor spoofing attack. The point number of $T$ can be 20, 40 and 60 to represent different attack capabilities as mentioned before. 
In the next section, the attack capability $\mathcal{A}$ modeled here is used to model the perturbation of the \featuremap{} $x$.

\subsection{Input Perturbation Modeling}

\label{subsec:input_perturb}

After analyzing spoofing attack capability $\mathcal{A}$, to formulate $x \oplus t'$ in Equation~\ref{eq:aml_goal_opt}, We need to have the following steps: 
(1) formulating the merging function $\oplus$;
(2) modeling the \advfeature{} spoofing capability $\Phi(\mathcal{A})$ based on known spoofing attack capability $\mathcal{A}$. 
In this section, we first formulate the merging function $\oplus$ by analyzing the pre-processing program. Then we model the \advfeature{} spoofing capability $\Phi(\mathcal{A})$ by expressing $t'$ with \advfeature{} $t$ in a differentiable function using global spatial transformations. Here, \advfeature{} $t$ can be attained with a given \advsensor{} $T$ by $t=\Phi(T)$.



\textbf{Formulating merging function ($\oplus$).}
To model the merging function $\oplus$ operated on $x$ and $t'$, which are in the domain of \featuremap{},
we need to first analyze the pre-processing program $\Phi(\cdot)$ that transforms the \sensor{} $X$ into the \featuremap{} $x$. As described in~\S\ref{subsec:LiDAR_object_detection}, the pre-processing process consists of three sub-processes: {\em coordinate transformation}, {\em ROI filtering} and {\em \featuremap{} extraction}. The first two processes make minor effects on the adversarial \advsensor{} $T'$ generated by the spoofing attack we conducted in~\S\ref{sec:input_perturb}. The {\em coordinate transformation} process has no effect because the adversarial \advsensor{} $T'$ will be transformed along with the \sensor{} X. As for the {\em ROI filtering} process, it filters out \pointcloud{} located outside of the road from a bird's-eye view. Therefore, as long as we spoof points on the road, the {\em ROI filtering process} makes no effect on the adversarial \advsensor{} $T'$.
The {\em feature extraction} process, as we mentioned in Section~\ref{subsec:LiDAR_object_detection}, extracts statistical features such as average height ($I_{avg\_h}$), average intensity ($I_{avg\_int}$), max height ($I_{max\_h}$) and so on. 

Because of such pre-processing, the \advfeature{} $t'$ cannot be directly added to the \featuremap{} $x$ to attain the adversarial \featuremap{} $x'$. To attain $x'$, we express such ``addition'' operation ($\oplus$) as a differentiable function shown below. Note that in this equation we do not include a few features in Table~\ref{table:cnn_input} such as direction and distance since they are either constant or can be derived directly from the features included in the equation.

\begin{equation}
\label{eq:phi_merge}
\begin{split}
    x' &= x \bigoplus \advdeltad \\
    &=
    \begin{bmatrix}
      &I^x_{cnt}+I^{\advdeltad}_{cnt}\\
                  &(I^x_{avg\_h}\cdot I^x_{cnt} + I^{\advdeltad}_{avg\_h} \cdot I^{\advdeltad}_{cnt})/(I^x_{cnt} + I^{\advdeltad}_{cnt})\\
                  &\max(I^x_{max\_h},I^{\advdeltad}_{max\_h})\\
                  &(I^x_{avg\_int}\cdot I^x_{cnt} + I^{\advdeltad}_{avg\_int} \cdot I^{\advdeltad}_{cnt})/(I^x_{cnt} + I^{\advdeltad}_{cnt})\\
                  & \sum I_{max\_int}^x 
                  \cdot  \mathbf{1}
                  \{ I_{max\_h}^x = 
                  max\{I_{max\_h}^x,
                  I_{max\_h}^{\advdeltad}\} \}
    \end{bmatrix}
\end{split}
\end{equation} 

\textbf{Modeling \featuremap{} spoofing capability $\Phi(\mathcal{A})$.}
To model \featuremap{} spoofing capability $\Phi{(\mathcal{A})}$, it equals to representing adversarial \featuremap{} $t'$ with known \advfeature{} $t$. 
We can use global spatial transformations including rotation, translation and scaling, under certain constraints to represent the \featuremap{} spoofing capability. Here the translation and scaling transformation interprets the attack capability in terms of modifying the \emph{azimuth} of \sensor{} while the rotation transformation interprets the attack capability in terms of modifying the \emph{distance} of \sensor{} from the LiDAR. 






Specifically, we apply the global spatial transformation to a set of the \advfeature{} $\mathcal{S}_\mathrm{t}$ to formulate the \advfeature{} spoofing capability $\Phi(\mathcal{A})$ and to represent adversarial \advfeature{} t'. 
For each \advfeature{} $t \in \mathcal{S}_\mathrm{t}$, it is mapped from a corresponding \advsensor{} $T$ such that $t = \Phi(T)$.

We use $t'_{(i)}$ to denote values of the $i$-th position  on the \advfeature{} $t'$ and 2D coordinate ($u^{'}_{(i)}, v^{'}_{(i)}$) to denote its location. $t'$ is transformed from an arbitrary instance $t$ where $t\in \mathcal{S}_\mathrm{t}$ by applying a homography matrix $H(\theta, \tau, \epsilon)$. 
The location of $t_{(i)}$ can be derived as $t'_{(i)}$ as follows:
\begin{equation}
\begin{split}
    (u_{(i)},v_{(i)}, 1)^{T}  = H \cdot (u^{'}_{(i)}, v^{'}_{(i)}, 1)^{T}, \\
    \mathbf{w.r.t. } \qquad H =
\begin{bmatrix}
  \epsilon (\cos{\theta} & -\sin{\theta}) & \tau_x\\
  \epsilon (\sin{\theta} & \cos{\theta}) & \tau_y\\
  0 & 0 & 1
\end{bmatrix}
\end{split}
\end{equation} 
Notice that here, $\tau_x/\tau_y$ has a fixed ratio $\tan\theta$ since the translation is performed along the $r$ axis shown in Fig.~\ref{fig:perturb_capabilities} (1). Since $\theta$ is dependent on the \advfeature{} we provide for performing the transformation, we align the \advfeature{} in advance to the $x$ axis where $\theta = 0$ and accordingly $\tau_y = \tau_x\tan\theta = 0$. Therefore, we can optimize $\tau_x$ alone. Also, this process is equivalent to scaling so we remove $\epsilon$.

We use the differentiable bilinear interpolation~\cite{jaderberg2015spatial} to calculate $t'_{(i)}$: 
\begin{equation}
\label{equ:bilinear}
t'_{(i)} = \sum_{q\in \mathcal{N}(u_{(i)}, v_{(i)})} t_{(q)} (1-|u_{(i)}- u_{(q)}|) (1 - |v_{(i)}-v_{(q)}|),
\end{equation}
where $\mathcal{N}(u_{(i)}, v_{(i)})$ represents the 4-pixel neighbors (top-left, top-right,bottom-left, bottom-right) at the location ($u(i), v(i)$) . 

Further, we can observe that the \featuremap{} contains the height information as shown in Table~\ref{table:cnn_input}. \cut{It can be further achieved by spoofing points at targeted vertical lines.}
So we also optimize a global scale scalar $s_h$ to the height features when generating adversarial \advfeature{} $t'$. Define $S(t, s_h)$ as the scaling function that multiplies the features which contain the height information by $s_h$. Based on this transformation, Equation~\ref{equ:bilinear} will be changed as follows. For simplification, we denote the whole transformation progress as $G_t$. So $G_t(\theta, \tau_x, s_h; t)$ represents the transformed adversarial \advfeature given \advfeature{} $t$ with transformation parameters $\theta, \tau_x, s_h$. 
\begin{equation}\label{eq:adv-delta}
\begin{split}
t'_{(i)} &= G_{t(i)}(\theta, \tau_x, s_h; t)\\ & =  \sum_{q\in \mathcal{N}(u_{(i)}, v_{(i)})}  S(t_{(q)}, s_h) (1-|u_{(i)}- u_{(q)}|) (1 - |v_{(i)}-v_{(q)}|)
\end{split}
\end{equation}

\section{Generating Adversarial Examples}
\label{sec:aml_analysis}

\begin{figure*}[t]
  \centering
    \includegraphics[width=1.0\textwidth]{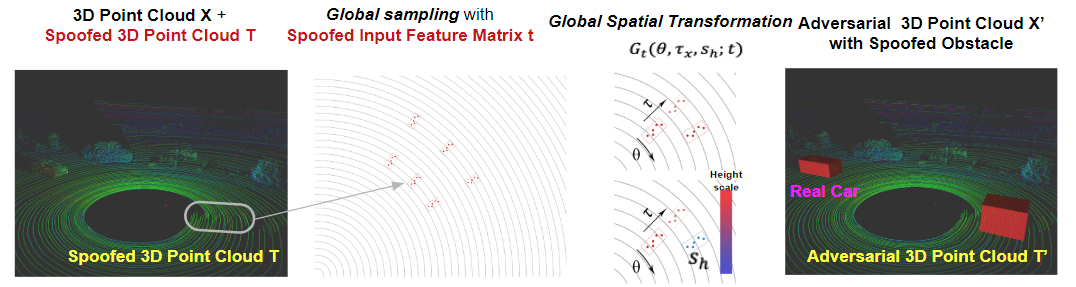}
    \caption{Overview of the adversarial example generation process.}
    \label{fig:gen-ae}
\end{figure*}

After modeling the input perturbation, in this section we design the objective function with an effective adversarial loss $\gL_{\adv}$, and leverage an optimization method to find the attack transformation parameters that minimize such loss.


\textbf{Design the adversarial loss $\gL_{\adv}$.}
Unlike previous work that performs the analysis only at the machine learning model level, there is no obvious objective function reflecting our attack goal of spoofing front-near obstacles. Yet, creating an effective objective function has been shown to be essential in generating effective adversarial examples~\cite{carlini2017towards}. In order to design an effective objective function, we analyze the post-processing step for the machine learning output. As shown in~\S\ref{subsec:LiDAR_object_detection}, in the clustering process, each cell of the model output is filtered by its \emph{objectness} value. After the clustering process, candidate object clusters are filtered by their \emph{positiveness} values. Upon such observation, we designed the adversarial loss $\gL_{\adv}$ as follows,
\begin{equation}\label{eq:adv_loss}
\resizebox{\linewidth}{!}{$
    \gL_{adv} = \sum (1 - Q( x', \mathrm{positiveness})  Q( x', \mathrm{objectness})) \mathcal{M}(px, py)$
    }
\end{equation}
where $Q(x', \cdot)$ is the function to extract the probabilities of $\cdot$ attribute from model $M$ by feeding in adversarial example $x'$.  $\mathcal{M}$ is a standard Gaussian mask with center coordinate $(px,py)$ which is an attack target position chosen by the attacker. We attain $(px,py)$ by mapping the attack target position in the real world onto the corresponding coordinates of the cell in the \featuremap{} using $\Phi$. The adversarial loss is then the summation over all the cells in the \featuremap{} of the weighted value described above. By minimizing this designed adversarial loss, it equals to increasing the probability to detect the obstacle of the adversarial \advsensor{} given the machine learning model $M$. 

\textbf{Optimization algorithm and our improvement using sampling.} With the $\gL_{\adv}$ design above, the optimization problem can be directly solved by using the Adam optimizer~\cite{kingma2014adam} to obtain the transformation parameters $\theta, \tau_x$ and scalar $s_h$ by minimizing the following objective function: 
\begin{equation}\label{eq:adv}
\resizebox{\linewidth}{!}{$
    f = \argmin_{\theta, \tau_x, s_h} \sum (1 - Q(x', \mathrm{positiveness})  Q( x', \mathrm{objectness})) \mathcal{M}(px, py)$
    }
\end{equation}
where $t'$ can be obtained by Equation~\ref{eq:adv-delta} and $x' = x\oplus t'$. In this paper, we call this direct solution \textit{vanilla optimization}. 


\begin{figure}[t]
  \centering
    \includegraphics[width=0.5\textwidth]{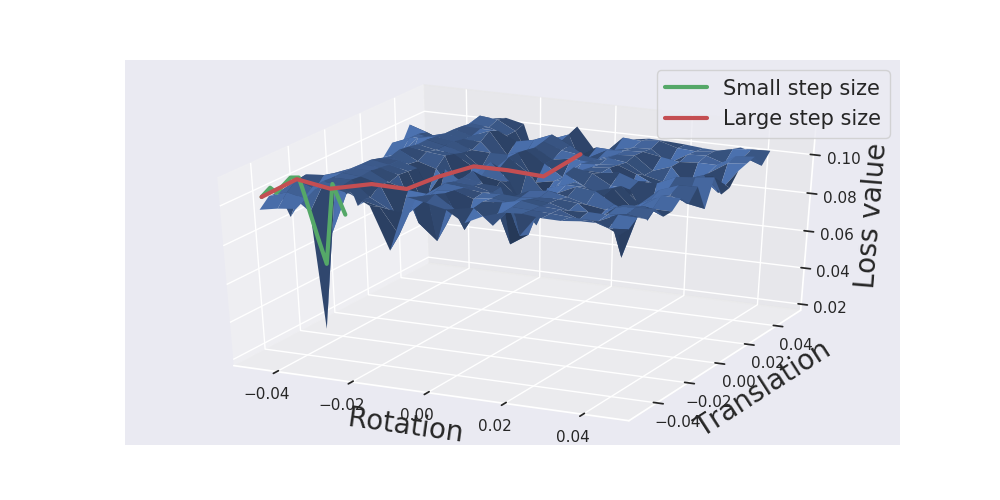}
    \caption{Loss surface over transformation parameters $\theta$ (rotation) and $\tau_x$ (translation). Using a small step size (green line) will trap the optimizing process near a local extreme while choosing a large step size (red line) will be less effective.}
    \label{fig:loss_surface}
\end{figure}

We visualize the loss surface against the transformation parameters in Fig.~\ref{fig:loss_surface}. During the vanilla optimization process, we observe that the loss surface over the transformation parameters is noisy at a small scale (green line) and quite flat at a large scale (red line). This leads to the problem of choosing a proper step size for optimization-based methods. For example, choosing a small step size will trap the optimizing process near a local minimum while choosing a large step size will be less effective due to noisy local loss pointing to the wrong direction. Different from Carlini et al.~\cite{carlini2017towards} that directly chose multiple starting points to reduces the trap of local minima, the optimization process in our setting is easy to get stuck in bad local minima due to the hard constraints of the perturbations.  
 We propose a way to use {\em sampling} at a larger scale and to {\em optimize} at a smaller scale. 
To initiate the optimization process at different positions, we first calculate the range of the transformation parameters so that the transformed \advsensor{} is located in the target area. Then we uniformly take $n$ samples for rotation and translation parameters and create $n^2$ samples to initiate with. 

\textbf{Generating adversarial \advsensor{}.}
To further construct the adversarial \sensor{} $X'$, we need to construct adversarial \advsensor{} $T'$. Using the transformation parameters $\theta, \tau, \epsilon, s_h$, we can express the corresponding adversarial \advsensor{} $T'$ such that $t' = \Phi(T')$ with a dual transformation function $G_T$ of $G_t$. We use $T_{\wx}, T_{\wy}, T_{\wz}$ to denote value of coordinate ($\wx,\wy,\wz$) and $T_i$ to denote the value of intensity for all points in \advsensor{} $T$. With transformation parameters $\theta, \tau, \epsilon, s_h$, we can express $T_{\wx}', T_{\wy}', T_{\wz}'$ of the transformed adversarial \advsensor{} $T'$ in Equation~\ref{eq:adv-sensor-delta}.

\begin{equation}\label{eq:adv-sensor-delta}
\begin{split}
& \quad T_i' = T_i \\
\begin{bmatrix}
  T_{\wx}'\\
  T_{\wy}'\\
  T_{\wz}'\\
  1 
\end{bmatrix} & = \begin{bmatrix}
  \cos{\theta} & -\sin{\theta} & 0 & \tau_x\\
  \sin{\theta} & \cos{\theta} & 0 & 0\\
  0 & 0 & s_h & 0 \\
  0 & 0 & 0 & 1 \\
\end{bmatrix} 
\cdot 
\begin{bmatrix}
  T_{\wx}\\
  T_{\wy}\\
  T_{\wz}\\
  1
\end{bmatrix}
\end{split}
\end{equation}

Therefore, we can use $T' = G_T(\theta,\tau_x,s_h;T)$ represents the transformed adversarial \advsensor{} given \advsensor{} $T$ with transformation parameters $\theta, \tau_x, s_h$. 

\textbf{Overall adversarial example generation process}. 
Fig.~\ref{fig:gen-ae} provides an overview of the overall adversarial example generation process. Given \sensor{} X and \advsensor{} $T$ (Fig.~\ref{fig:gen-ae} (a)), we first map them via $\Phi$ to get corresponding \featuremap{} $x$ and \advfeature{} $t$. Then we apply the sampling algorithm to initialize the transformation parameters $\theta,\tau_x,s_h$ as shown in Fig.~\ref{fig:gen-ae} (b).  After the initialization, we leverage optimizer $opt$ to further optimize the transformation parameters ($\theta,\tau_x, s_h$) with respect to the adversarial loss function $\gL_{\adv}$ (Fig.~\ref{fig:gen-ae} (c)). With the transformation parameters $\theta,\tau_x,s_h$ and $T$, we apply the dual transformation function $G_T$ using the Equation~\ref{eq:adv-sensor-delta} to get adversarial \advsensor{} $T'$. At last, to obtain the adversarial \sensor{} $X'$, we append $T'$ to \sensor{} $X$ (Fig.~\ref{fig:gen-ae} (d)).
The entire adversarial example generation algorithm including the optimization parameters is detailed in Appendix~\ref{appendix:algorithm}.
\section{Evaluation and Results}
\label{sec:eval}
In this section, we evaluate our adversarial example generation method in terms of attack effectiveness and robustness.

\textbf{Experiment Setup.} We use the real-world LiDAR sensor data trace released by Baidu Apollo team with Velodyne HDL-64E S3, which is collected for 30 seconds on local roads at Sunnyvale, CA. We uniformly sample 300 \sensor{} frames from this trace in our evaluation. The attack goal is set as spoofing an obstacle that is 2-8 meters to the front of the victim AV. The distance is measured from the front end of the victim AV to the rear end of the obstacle.

\subsection{Attack Effectiveness}
\label{sec:single-eval}
\begin{figure}[t]
  \centering
    \includegraphics[width=0.45\textwidth]{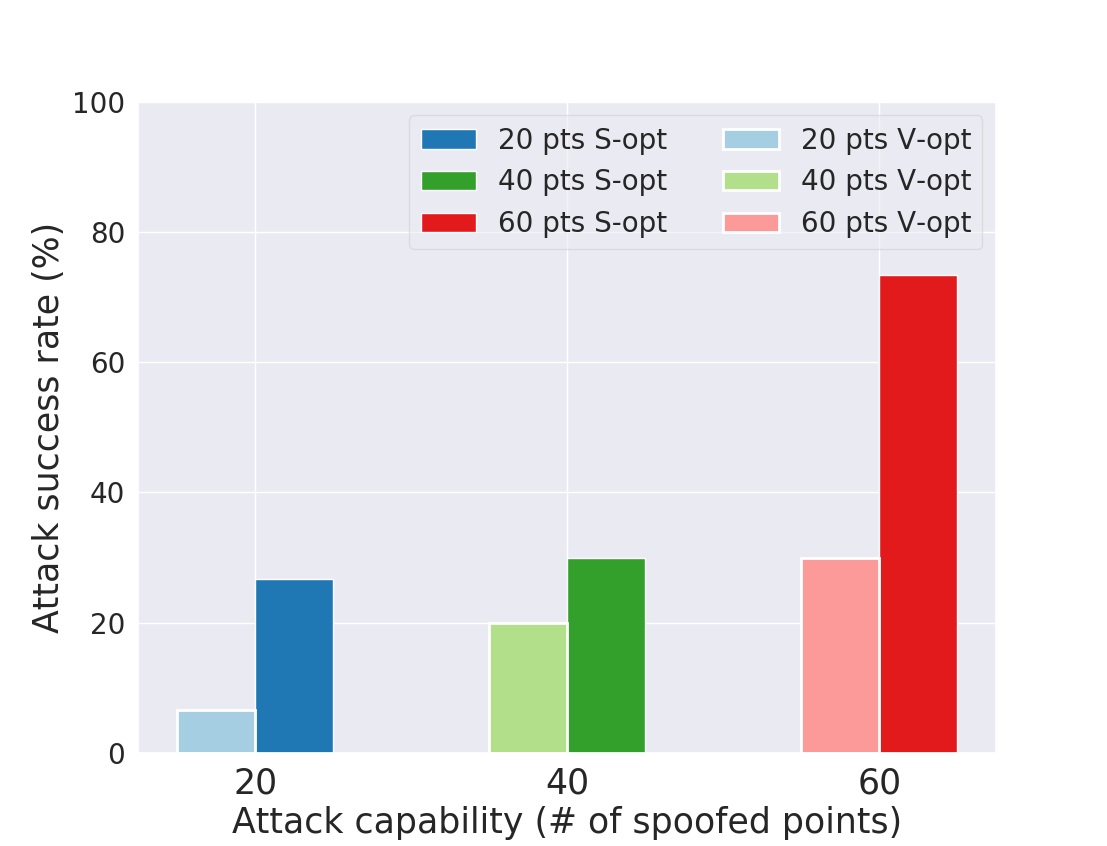}
    \caption{Attack success rate of spoofing a front-near obstacle with different number of spoofed points. V-opt refers to vanilla optimization which is directly using the optimizer and S-opt refers to sampling based optimization. We choose Adam~\cite{kingma2014adam} as the optimizer in both cases.}
    \label{fig:single-eval}
\end{figure}

Fig.~\ref{fig:single-eval} shows the success rates of generating a spoofed obstacle with different attack capabilities using the vanilla optimization and our improved optimization with global sampling (detailed in~\S\ref{sec:aml_analysis}). As shown, with our improvement using sampling, the success rates of spoofing front-near obstacles are increased from 18.9\% to 43.3\% on average, which is a 2.65$\times$ improvement. This shows that combining global sampling with optimization is effective in addressing the problem of becoming trapped in local minima described in~\S\ref{sec:aml_analysis}.

Fig.~\ref{fig:single-eval} also shows that the success rates increase with more spoofed points, which is expected since the attack capability is increased with more spoofed points. In particular, when the attacker can reliably inject 60 spoofed points, which is the attack capability observed in our experiments (\S\ref{sec:spoofing-attack}), the attack is able to achieve around $75\%$ success rate using our improved optimization method.

In addition, we observe that the spoofed obstacles in all of the successful attacks are classified as vehicles after the LiDAR-based perception process, even though we do not specifically aim at spoofing vehicle-type obstacles in our problem formulation.

\subsection{Robustness Analysis}

In this section, we perform analysis to understand the robustness of the generated adversarial \advsensor{} $T'$ to variations in \sensor{} $X$ and \advsensor{} $T \in \mathcal{S}_\mathrm{T}$. Such analysis is meaningful for generating adversarial \advsensor{} that has high attack success rate in the real world. To launch the attack in the real world, there are two main variations that affect the results: variation in spoofed points and variation in positions of the victim AV. 1) The imprecision in the attack devices contributes to the variation of the spoofed points. The attacker is able to stably spoof 60 points at a global position as we state in ~\S\ref{sec:lidar-spoofing}. However, it is difficult to spoof points with precise positions. It is important to understand whether such imprecision affects the attack success rate. 2) The position of the victim AV is not controlled by the attacker and might vary from where the attacker collected the \sensor{}. It is important to understand whether such difference affects the attack success rate.

\begin{figure}
  \centering
    \includegraphics[width=0.45\textwidth]{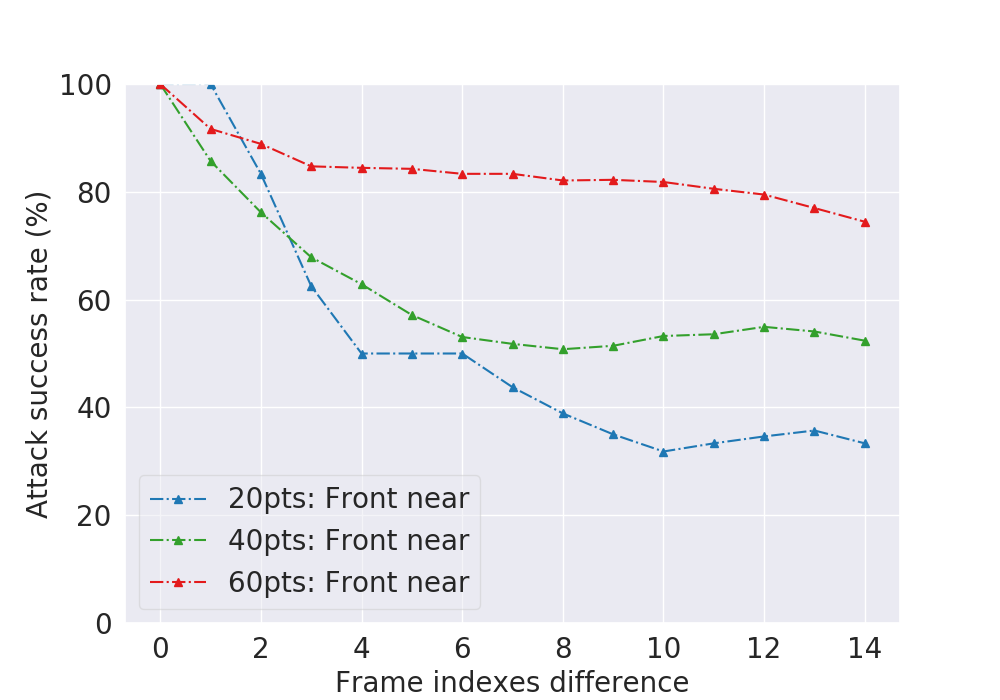}
    \caption{The robustness of the generated adversarial \advsensor{} to variations in \sensor{} $X$. We quantify the variation in \sensor{} $X$ as the frame indexes difference between the evaluated \sensor{} and the \sensor{} used for generating the adversarial \advsensor{}.}
    \label{fig:robust-X}
\end{figure}

\textbf{Robustness to variations in point cloud.} To measure the robustness to variations in the \sensor{}, we first select all the \sensor{} frames that can generate successful adversarial \advsensor{}. For each of them, we apply its generated adversarial \advsensor{} to 15 consecutive frames (around 1.5 s) after it and calculate the success rates. Fig.~\ref{fig:robust-X} shows the analysis results. In this figure, the x-axis is the index for the 15 consecutive frames, and thus the larger the frame index is, the larger the variation is from the original \sensor{} that generates the adversarial \advsensor{}. As shown, the robustness for attacks with more spoofed points is generally higher than that for attacks with fewer spoofed points, which shows that higher attack capability can increase the robustness. Particularly, with 60 spoofed points, the success rates are on average above 75\% during the 15 subsequent frames, which demonstrates a high degree of robustness. This suggests that launching such attack does not necessarily require the victim AV to be located at the exact position that generates the adversarial example in order to have high success rates. 


\textbf{Robustness to variations in \advsensor{}.}
To evaluate the robustness to variations in the \advsensor{}, for a given \advsensor{} $T \in \mathcal{S}_\mathrm{T}$, we first generate the corresponding adversarial \advsensor{} $T'$ with a \sensor{} $X$. Next, we generate 5 more \advsensor{} traces $T_1, ..., T_5$ using our LiDAR spoofing attack experiment setup. Next, we use the same transformation that generates $T'$ from $T$ to generate $T'_1, ..., T'_5$, and then combine each of them with $X$ to launch the attack. Table~\ref{table:attack-trace-eval} shows the average success rates with different attack capabilities. As shown, for all three attack capabilities we are able to achieve over 82\% success rates. With 60 spoofed points, the success rate is as high as 90\%. This suggests that launching such attack does not require the LiDAR spoofing attack to be precise all the time in order to achieve high success rates.



\begin{table}[t]
\centering
\begin{tabular}{ |p{1.5cm}|p{0.5cm}|p{1.0cm}|p{1.0cm}|p{1.0cm}|  }
 \hline
  \multicolumn{1}{|c|}{Targeted position}  & \multicolumn{3}{|c|}{\# Spoofed points} \\
  \cline{2-4}
  \multicolumn{1}{|c|}{}& 20 & 40 & 60\\
 \hline
  2-8 meters     & 87\%    &82\%  &   90\% \\
 \hline
\end{tabular}
\centering
 \vspace{0.1in}
     \caption{Robustness analysis results of generated adversarial \advsensor{} to variation in \advsensor{} $T \in \mathcal{S}_\mathrm{T}$. The robustness is measured by average attack success rates. 
     }
     \label{table:attack-trace-eval}
     \vspace{-0.2in}
\end{table}
\section{Driving Decision Case Study}
\label{sec:e2e}

To understand the impact of our attacks at the driving decision level, in this section we construct two attack scenarios and evaluate them on Baidu Apollo using their simulation features as case studies.



\textbf{Experiment setup.} We perform the case study using the simulation feature provided by Baidu Apollo, called \textit{Sim-control}, which is designed to allow users to observe the AV system behavior at the driving decision level by replaying collected real-world sensor data traces. Sim-control does not consist of a physics engine to simulate the control of the vehicle. Instead, the AV behaves exactly the same as what it plans. Although it cannot directly reflect the attack consequences in the physical world, it can serve for our purpose of understanding the impact of our attacks on AV driving decisions.

For each attack scenario in the case study, we simulate it in Sim-control using synthesized continuous frames of successful adversarial \sensor{} identified in \S~\ref{sec:eval} as input. The experiments are performed on Baidu Apollo 3.0.





\begin{figure*}[!ht]
  \begin{minipage}{0.64\textwidth}
\includegraphics[width=0.99\textwidth]{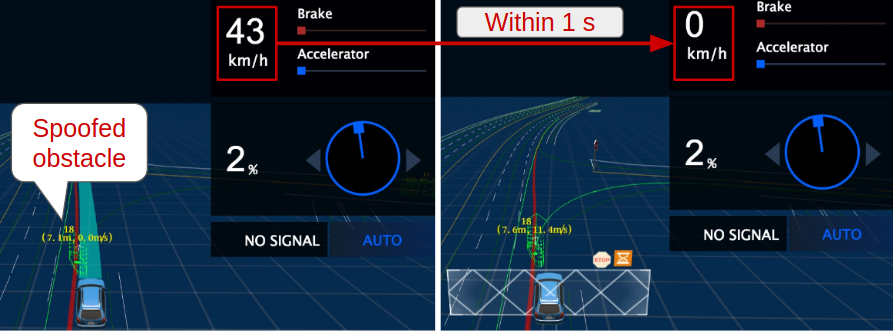}
    \captionof{figure}{Demonstration of the emergency brake attack. Due to the spoofed obstacle, the victim AV makes a sudden stop decision to drop its speed from 43 km/h to 0 km/h within a second, which may cause injuries of passengers or rear-end collisions.}
    \label{fig:hard-brake}
\end{minipage}
  \hspace{0.03in}
   \begin{minipage}{0.34\textwidth}
   \hspace{0.02in}
    \includegraphics[width=0.92\textwidth]{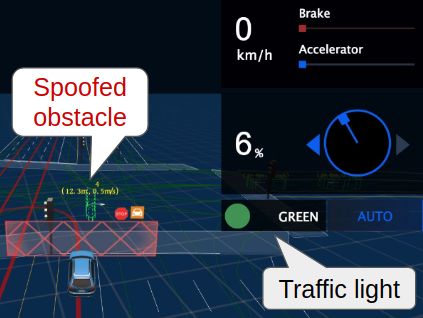}
    \captionof{figure}{Demonstration of the AV freezing attack. The traffic light is turned green but the victim AV is not moving due to the spoofed front-near obstacles.}
    \label{fig:tl_case}
\end{minipage}
\end{figure*}

\textbf{Case study results.} We construct and evaluate two attack scenarios in this case study\footnote{Video demos can be found at \url{http://tinyurl.com/advlidar}}:


{(1) Emergency brake attack.} In this attack, we generate adversarial \sensor{} that spoofs a front-near obstacle to a moving victim AV. We find that the AV makes a stop decision upon this attack.
As illustrated in Fig.~\ref{fig:hard-brake}, the stop decision triggered by a spoofed front-near obstacle causes the victim AV to decrease its speed from 43 km/h to 0 km/h within 1 second. This stop decision will lead to a hard brake~\cite{hard_brake}, which may hurt the passengers or result in rear-end collisions. Noticed that, Apollo does implement driving decisions for overtaking. However, for overtaking, a minimum distance is required based on the relative speed of the obstacle. Therefore, with our near spoofed obstacle, the victim AV makes stop decisions instead of overtaking decisions.



{(2) AV freezing attack.} In this attack, we generate an adversarial \sensor{} that spoofs an obstacle in front of an AV victim when it is waiting at a red traffic light. We simulate this scenario with the data trace at an intersection with a traffic light. As shown in Fig.~\ref{fig:tl_case}, since the victim AV is static, the attacker can constantly attack and prevent it from moving even after the traffic signal turns green, which may be exploited to cause traffic jams. Noticed that, Apollo does implement driving decisions for deviating static obstacles. However, for deviation or side passing, it requires a minimum distance (15 meters by default). Therefore, with our near spoofed obstacle, the victim AV makes stop decisions instead of side passing decisions.

\section{Discussion}
\label{sec:discussion}
In this section, we discuss the limitations and generality of this study. We then discuss potential defense directions.
\subsection{Limitations and Future Work}
\textbf{Limitations in the sensor attack.}
One major limitation is that our current results cannot directly demonstrate attack performance and practicality in the real world. For example, performing our attack on a real AV on the road requires dynamically aiming an attack device at the LiDAR on a victim car with high precision, which is difficult to prove the feasibility without road tests in the physical world.
In this work, 
our goal is to 
provide new understandings of this research problem.
Future research directions include conducting real world testing. To demonstrate the attack in the real world, we plan to first conduct the sensor attack with LiDAR on top of a real vehicle in outdoor settings. In this setting, the sensor attack could be enhanced by: 1) enlarging the laser spoofing area to solve the aiming problem; 2) adjusting the delay time so that the attacker could spoof points at different angles without moving the attack devices.
Then we could apply our proposed methodology to conduct drive-by experiments in different attack scenarios mentioned in ~\S\ref{sec:e2e}. 

\textbf{Limitations in adversarial example generation.}
First, we construct adversarial sensor data by using a subset of spoofing attack capability $\mathcal{A}$. Therefore, our analysis may not fully reveal the full potential of sensor attacks.
Second, though we have performed the driving decision case study, we did not perform a comprehensive analysis on modules beyond the perception module. That means that the designed objective function can be further improved to more directly target specific abnormal AV driving decisions.
\subsection{Generality on LiDAR-based AV Perception}

\textbf{Generality of the methodology.} Attacking any LiDAR-based AV perception system with an adversarial sensor attack can be formulated as three components: (1) formulating the \advsensor{} capability $\mathcal{A}$, (2) generating adversarial examples, and (3) evaluating at the driving decision level. Even though our construction of these components might be specific to Baidu Apollo, our analysis methodology can be generalized to other LiDAR-based AV perception systems.

\textbf{Generality of the results.} The formulation of \sensor{} spoofing capability $\mathcal{A}$ can be generalized as it is independent from AV systems. 
The success of the attack may be extended to other LiDAR-based AV perception system due to the nature of the LiDAR sensor attack. The LiDAR spoofing attack introduces a \advsensor{}, which was not foreseen in the training process of machine learning models used in the AV perception system. Therefore, other models are also likely to be vulnerable to such spoofing patterns.


\subsection{Defense Discussion}
This section discusses defense directions at AV system, sensor, and machine learning model levels.
\subsubsection{AV System-Level defenses}
In our proposed attack, the attacker only needs to inject at most 60 points to spoof an obstacle, but the \sensor{} of a detected real vehicle can have as many as a thousand points (can be illustrated in Fig.~\ref{fig:side-compare}). We look into the point cloud of a detected spoofed obstacle and find that the \sensor{} consists of points reflected from the ground, in addition to the points spoofed by the attacker. For example, one of the successful adversarial \advsensor{} we generated with 20 spoofed points is detected as an obstacle containing 283 points.

Points from ground reflection are clustered into obstacles due to the information loss introduced in the pre-processing phase. More specifically, mapping a 3D point cloud into a 2D matrix results in height information loss. This vulnerability contributes to the success of the proposed attack. To mitigate the impacts of this problem, we propose two defenses at the AV system level: (1) filtering out the ground reflection in the pre-processing phase, and (2) either avoiding transforming \sensor{} into \featuremap{} or adding more features to reduce the information loss.

\subsubsection{Sensor-Level Defenses}
Several defenses could be adopted against spoofing attacks on LiDAR sensors:


\textbf{Detection techniques.} \cut{A defense based on sensor fusion consists of combining data from multiple sensors to detect anomalies. Advanced driver-assistance systems often include other sensors beyond LiDAR. Radars, ultrasonic sensors, and cameras provide additional information and sensing redundancy to the controller that can be used to detect and handle an attack. For instance, radars could be used to verify if the perception of the LiDAR is consistent. }
Sensor fusion, which intelligently combines data from several sensors to detect anomalies and improve performance, could be adopted against LiDAR spoofing attacks. AV systems are often equipped with sensors beyond LiDAR. Cameras, radars, and ultrasonic sensors provide additional information and redundancy to detect and handle an attack on LiDAR.

Different sensor fusion algorithms have been proposed focusing on the security and safety aspects \cite{yang2018sensor} \cite{ivanov2014attack}. However, the sensor fusion defense requires the majority of sensors to be functioning correctly. While not a perfect defense, sensor fusion approaches can significantly increase the effort of an attacker.


\textbf{Mitigation techniques.} Another class of defenses aims to reduce the influence of the attack by modifying the internal sensing structure of the LiDAR. Different solutions include reducing the receiving angle and filtering unwanted light spectra to make LiDARs less susceptible to attacks \cite{sensor-blackhat15, Shin2017IllusionAD}. However, these techniques also reduce the capacity of the LiDAR to measure the reflected laser pulses, which limits the range and the sensitivity of the device.

\textbf{Randomization techniques.} Another defense is adding randomness to how the LiDAR fires laser pulses. The attacker cannot know when to and what laser pulses to fire if the LiDAR fires laser pulses with an unpredictable pattern. A solution could be firing a random grouping of laser pulses each cycle. An attacker would not know which reflections the LiDAR would be expecting.
Another alternative would be randomizing the laser pulses waveform. With sensitive equipment, it would be possible to only accept reflection waveforms that match randomized patterns uniquely produced by the LiDAR laser.  
Another solution, proposed by Shoukry et al. \cite{Shoukry2015PyCRAPC}, consists of randomly turning off the transmitter to verify with the receiver if there are any unexpected incoming signals.
Adding randomness makes it difficult for an attacker to influence the measurements, but this approach also adds significant complexity to the overall system and trades off with performance.

\subsubsection{Machine Learning Model-Level Defense}
Various detection and defense methods have also been explored~\cite{ma2018characterizing,madry2017towards,carlini2017adversarial,athalye2018obfuscated} against adversarial examples in image classification.  Adversarial training ~\cite{goodfellow2014explaining} and its variations~\cite{tramer2017ensemble,madry2017towards} are more successful to improve the robustness of the model. 
Motivated by the adversarial examples generated by our algorithm, we can combine them with the original training data to conduct adversarial retraining and thus improve the model robustness. \cut{\chaowei{maybe we do not need the following sentence }
However, Recently, \cite{athalye2018obfuscated} successfully generated adversarial examples in the presence of detection and defense strategies. So the defense against adversarial examples is still an open problem in adversarial machine learning domain. }



\section{Related Work}
\textbf{Vehicle systems security.} 
Numerous previous works explore security problems in vehicle systems and have uncovered vulnerabilities in in-vehicle networks of modern automobiles~\cite{vehicle-oakland10,vehicle-sec11,fault-vehicle-ccs16}, infotainment systems~\cite{mazloom2016security}, and emerging connected vehicle-based systems~\cite{cvattack-ndss18, trb:2018:yiheng:signalsecurity, wong2019trajectory}. In comparison, our work focuses on vehicle systems with the emerging autonomous driving technology and specifically targets the security of LiDAR-based AV perception, which is an attack surface not presented in traditional vehicle systems designed for human drivers. 

\textbf{Vehicle-related sensor attacks.} 
The sensors commonly used in traditional vehicles have been shown to be vulnerable to attacks. Rouf et al. showed that tire pressure sensors are vulnerable to wireless jamming and spoofing attacks~\cite{RoufTPMSAttack}. Shoukry et al. attacked the anti-lock braking system of a vehicle by spoofing the magnetic wheel speed sensor~\cite{ShoukryABSSpoofing}. As AVs become popular, so have attacks against their perception sensors. Yan et al. used spoofing and jamming attacks to attack the ultrasonic sensors, radar, and camera on a Tesla Model S~\cite{Yan2016CanYT}. There have also been two works exploring the vulnerability of LiDAR to spoofing and jamming attacks~\cite{sensor-blackhat15, Shin2017IllusionAD}. In this work, we build on these prior work to show that LiDAR spoofing attacks can be used to attack the machine learning models used for LiDAR-based AV perception and affect the driving decision.

\textbf{Adversarial example generation.}
Adversarial examples have been heavily explored in the image domain~\cite{goodfellow2014explaining,xiao2018spatially,carlini2017towards,papernot2017practical}. \citet{xie2017adversarial} generated adversarial examples for segmentation and object detection while \citet{cisse2017houdini} for segmentation and human pose estimation. Researchers also apply adversarial examples to the physical world to fool machine learning models~\cite{evtimov2017robust,eykholt2018physical,athalye2017synthesizing}. Compared to these previous work exploring adversarial examples in the image domain, this work explores adversarial examples for LiDAR-based perception. An ongoing work~\cite{xiang2018generating} studies the generation of 3D adversarial point clouds. However, such attack focuses on the digital domain and can not be directly applied to the context of AV systems. In comparison, our method is motivated to generate adversarial examples based on the capability of sensor attacks to fool the LiDAR-based perception models in AV systems. 


\section{Conclusion}
\label{sec:conclusion}
In this work, we perform the first security study of LiDAR-based perception in AV systems. We consider LiDAR spoofing attacks as the threat model, and set the attack goal as spoofing front-near obstacles. We first reproduce the state-of-the-art LiDAR spoofing attack, and find that blindly applying it is insufficient to achieve the attack goal due to the machine learning-based object detection process. We thus perform analysis to fool the machine learning model by formulating the attack task as an optimization problem. We first construct the input perturbation function using local attack experiments and global spatial transformation-based modeling, and then construct the objective function by studying the post-processing process. We also identify the inherent limitations of directly using optimization-based methods and design a new algorithm that increases the attack success rates by 2.65$\times$ on average. As a case study, we further construct and evaluate two attack scenarios that may compromise AV safety and mobility. We also discuss defense directions at AV system, sensor, and machine learning model levels.


\begin{acks}
We would like to thank Shengtuo Hu, Jiwon Joung, Jiachen Sun, Yunhan Jack Jia, Yuru Shao,
Yikai Lin, David Ke Hong, the anonymous reviewers, and our shepherd Zhe Zhou for providing valuable feedback on our work. This research was supported in part by an award from Mcity at University of Michigan, by the National Science Foundation under grants CNS-1850533, CNS-1330142, CNS-1526455 and CCF-1628991, by ONR under N00014-18-1-2020.
\end{acks}

%
\bibliographystyle{ACM-Reference-Format}

\bibliography{ccs-sample}


\begin{thebibliography}{00}


\ifx \showCODEN    \undefined \def \showCODEN     #1{\unskip}     \fi
\ifx \showDOI      \undefined \def \showDOI       #1{#1}\fi
\ifx \showISBNx    \undefined \def \showISBNx     #1{\unskip}     \fi
\ifx \showISBNxiii \undefined \def \showISBNxiii  #1{\unskip}     \fi
\ifx \showISSN     \undefined \def \showISSN      #1{\unskip}     \fi
\ifx \showLCCN     \undefined \def \showLCCN      #1{\unskip}     \fi
\ifx \shownote     \undefined \def \shownote      #1{#1}          \fi
\ifx \showarticletitle \undefined \def \showarticletitle #1{#1}   \fi
\ifx \showURL      \undefined \def \showURL       {\relax}        \fi
\providecommand\bibfield[2]{#2}
\providecommand\bibinfo[2]{#2}
\providecommand\natexlab[1]{#1}
\providecommand\showeprint[2][]{arXiv:#2}

\bibitem[\protect\citeauthoryear{??}{har}{2005}]%
        {hard_brake}
 \bibinfo{year}{2005}\natexlab{}.
\newblock \bibinfo{title}{{HARD BRAKE \& HARD ACCELERATION}}.
\newblock
  \bibinfo{howpublished}{\url{http://tracknet.accountsupport.com/wp-content/uploads/Verizon/Hard-Brake-Hard-Acceleration.pdf}}.
    (\bibinfo{year}{2005}).
\newblock


\bibitem[\protect\citeauthoryear{??}{arb}{2016}]%
        {arbexpress}
 \bibinfo{year}{2016}\natexlab{}.
\newblock \bibinfo{title}{{ArbExpress}}.
\newblock
  \bibinfo{howpublished}{\url{https://www.tek.com/signal-generator/afg2021-software-0}}.
    (\bibinfo{year}{2016}).
\newblock


\bibitem[\protect\citeauthoryear{??}{lid}{2017a}]%
        {lidar_intro}
 \bibinfo{year}{2017}\natexlab{a}.
\newblock \bibinfo{title}{{An Introduction to LIDAR: The Key Self-Driving Car
  Sensor}}.
\newblock
  \bibinfo{howpublished}{\url{https://news.voyage.auto/an-introduction-to-lidar-the-key-self-driving-car-sensor-a7e405590cff}}.
    (\bibinfo{year}{2017}).
\newblock


\bibitem[\protect\citeauthoryear{??}{apo}{2017}]%
        {apollo}
 \bibinfo{year}{2017}\natexlab{}.
\newblock \bibinfo{title}{{Baidu Apollo}}.
\newblock \bibinfo{howpublished}{\url{http://apollo.auto}}.
  (\bibinfo{year}{2017}).
\newblock


\bibitem[\protect\citeauthoryear{??}{lid}{2017b}]%
        {lidar_waymo}
 \bibinfo{year}{2017}\natexlab{b}.
\newblock \bibinfo{title}{{Google's Waymo Invests in LIDAR Technology, Cuts
  Costs by 90 Percent}}.
\newblock
  \bibinfo{howpublished}{\url{https://arstechnica.com/cars/2017/01/googles-waymo-invests-in-lidar-technology-cuts-costs-by-90-percent/}}.
    (\bibinfo{year}{2017}).
\newblock


\bibitem[\protect\citeauthoryear{??}{kit}{2017}]%
        {kitti_3d}
 \bibinfo{year}{2017}\natexlab{}.
\newblock \bibinfo{title}{{KITTI Vision Benchmark: 3D Object Detection}}.
\newblock
  \bibinfo{howpublished}{\url{http://www.cvlibs.net/datasets/kitti/eval_object.php?obj_benchmark=3d}}.
    (\bibinfo{year}{2017}).
\newblock


\bibitem[\protect\citeauthoryear{??}{lid}{2017c}]%
        {lidar_gm}
 \bibinfo{year}{2017}\natexlab{c}.
\newblock \bibinfo{title}{{What it Was Like to Ride in GM's New Self-Driving
  Cruise Car}}.
\newblock
  \bibinfo{howpublished}{\url{https://www.recode.net/2017/11/29/16712572/general-motors-gm-new-self-driving-autonomous-cruise-car-future}}.
    (\bibinfo{year}{2017}).
\newblock


\bibitem[\protect\citeauthoryear{??}{bai}{2018a}]%
        {baidu_volvo_ford}
 \bibinfo{year}{2018}\natexlab{a}.
\newblock \bibinfo{title}{{Baidu hits the gas on autonomous vehicles with Volvo
  and Ford deals}}.
\newblock
  \bibinfo{howpublished}{\url{https://techcrunch.com/2018/11/01/baidu-volvo-ford-autonomous-driving/}}.
    (\bibinfo{year}{2018}).
\newblock


\bibitem[\protect\citeauthoryear{??}{bai}{2018b}]%
        {baidu_apolong}
 \bibinfo{year}{2018}\natexlab{b}.
\newblock \bibinfo{title}{{Baidu starts mass production of autonomous buses}}.
\newblock
  \bibinfo{howpublished}{\url{https://www.dw.com/en/baidu-starts-mass-production-of-autonomous-buses/a-44525629}}.
    (\bibinfo{year}{2018}).
\newblock


\bibitem[\protect\citeauthoryear{??}{vel}{2018}]%
        {veloview}
 \bibinfo{year}{2018}\natexlab{}.
\newblock \bibinfo{title}{{VeloView}}.
\newblock \bibinfo{howpublished}{\url{https://www.paraview.org/VeloView/}}.
  (\bibinfo{year}{2018}).
\newblock


\bibitem[\protect\citeauthoryear{??}{lid}{2018a}]%
        {lidar_volvo}
 \bibinfo{year}{2018}\natexlab{a}.
\newblock \bibinfo{title}{{Volvo Finds the LIDAR it Needs to Build Self-Driving
  Cars}}.
\newblock
  \bibinfo{howpublished}{\url{https://www.wired.com/story/volvo-self-driving-lidar-luminar/}}.
    (\bibinfo{year}{2018}).
\newblock


\bibitem[\protect\citeauthoryear{??}{way}{2018}]%
        {waymo_public}
 \bibinfo{year}{2018}\natexlab{}.
\newblock \bibinfo{title}{{Waymo's autonomous cars have driven 8 million miles
  on public roads}}.
\newblock
  \bibinfo{howpublished}{\url{https://www.theverge.com/2018/7/20/17595968/waymo-self-driving-cars-8-million-miles-testing}}.
    (\bibinfo{year}{2018}).
\newblock


\bibitem[\protect\citeauthoryear{??}{lid}{2018b}]%
        {lidar_importance}
 \bibinfo{year}{2018}\natexlab{b}.
\newblock \bibinfo{title}{{What Is LIDAR, Why Do Self-Driving Cars Need It, And
  Can It See Nerf Bullets?}}
\newblock
  \bibinfo{howpublished}{\url{https://www.wired.com/story/lidar-self-driving-cars-luminar-video/}}.
    (\bibinfo{year}{2018}).
\newblock


\bibitem[\protect\citeauthoryear{??}{lyf}{2018}]%
        {lyft_public}
 \bibinfo{year}{2018}\natexlab{}.
\newblock \bibinfo{title}{{You can take a ride in a self-driving Lyft during
  CES}}.
\newblock
  \bibinfo{howpublished}{\url{https://www.theverge.com/2018/1/2/16841090/lyft-aptiv-self-driving-car-ces-2018}}.
    (\bibinfo{year}{2018}).
\newblock


\bibitem[\protect\citeauthoryear{Abadi, Barham, Chen, Chen, Davis, Dean, Devin,
  Ghemawat, Irving, Isard, et~al\mbox{.}}{Abadi et~al\mbox{.}}{2016}]%
        {abadi2016tensorflow}
\bibfield{author}{\bibinfo{person}{Mart{\'\i}n Abadi}, \bibinfo{person}{Paul
  Barham}, \bibinfo{person}{Jianmin Chen}, \bibinfo{person}{Zhifeng Chen},
  \bibinfo{person}{Andy Davis}, \bibinfo{person}{Jeffrey Dean},
  \bibinfo{person}{Matthieu Devin}, \bibinfo{person}{Sanjay Ghemawat},
  \bibinfo{person}{Geoffrey Irving}, \bibinfo{person}{Michael Isard},
  {et~al\mbox{.}}} \bibinfo{year}{2016}\natexlab{}.
\newblock \showarticletitle{Tensorflow: a system for large-scale machine
  learning.}. In \bibinfo{booktitle}{{\em OSDI}}, Vol.~\bibinfo{volume}{16}.
  \bibinfo{pages}{265--283}.
\newblock


\bibitem[\protect\citeauthoryear{Athalye, Carlini, and Wagner}{Athalye
  et~al\mbox{.}}{2018}]%
        {athalye2018obfuscated}
\bibfield{author}{\bibinfo{person}{Anish Athalye}, \bibinfo{person}{Nicholas
  Carlini}, {and} \bibinfo{person}{David Wagner}.}
  \bibinfo{year}{2018}\natexlab{}.
\newblock \showarticletitle{Obfuscated gradients give a false sense of
  security: Circumventing defenses to adversarial examples}.
\newblock \bibinfo{journal}{{\em arXiv preprint arXiv:1802.00420\/}}
  (\bibinfo{year}{2018}).
\newblock


\bibitem[\protect\citeauthoryear{Athalye and Sutskever}{Athalye and
  Sutskever}{2018}]%
        {athalye2017synthesizing}
\bibfield{author}{\bibinfo{person}{Anish Athalye} {and} \bibinfo{person}{Ilya
  Sutskever}.} \bibinfo{year}{2018}\natexlab{}.
\newblock \showarticletitle{{Synthesizing Robust Adversarial Examples}}. In
  \bibinfo{booktitle}{{\em International Conference on Machine Learning
  (ICML)}}.
\newblock


\bibitem[\protect\citeauthoryear{Carlini, Mishra, Vaidya, Zhang, Sherr,
  Shields, Wagner, and Zhou}{Carlini et~al\mbox{.}}{2016}]%
        {carlini2016hidden}
\bibfield{author}{\bibinfo{person}{Nicholas Carlini}, \bibinfo{person}{Pratyush
  Mishra}, \bibinfo{person}{Tavish Vaidya}, \bibinfo{person}{Yuankai Zhang},
  \bibinfo{person}{Micah Sherr}, \bibinfo{person}{Clay Shields},
  \bibinfo{person}{David Wagner}, {and} \bibinfo{person}{Wenchao Zhou}.}
  \bibinfo{year}{2016}\natexlab{}.
\newblock \showarticletitle{{Hidden Voice Commands}}. In
  \bibinfo{booktitle}{{\em USENIX Security Symposium}}.
\newblock


\bibitem[\protect\citeauthoryear{Carlini and Wagner}{Carlini and
  Wagner}{2017a}]%
        {carlini2017adversarial}
\bibfield{author}{\bibinfo{person}{Nicholas Carlini} {and}
  \bibinfo{person}{David Wagner}.} \bibinfo{year}{2017}\natexlab{a}.
\newblock \showarticletitle{{Adversarial Examples are not Easily Detected:
  Bypassing Ten Detection Methods}}. In \bibinfo{booktitle}{{\em Proceedings of
  the 10th ACM Workshop on Artificial Intelligence and Security}}. ACM,
  \bibinfo{pages}{3--14}.
\newblock


\bibitem[\protect\citeauthoryear{Carlini and Wagner}{Carlini and
  Wagner}{2018}]%
        {carlini2018audio}
\bibfield{author}{\bibinfo{person}{Nicholas Carlini} {and}
  \bibinfo{person}{David Wagner}.} \bibinfo{year}{2018}\natexlab{}.
\newblock \showarticletitle{{Audio Adversarial Examples: Targeted Attacks on
  Speech-to-text}}. In \bibinfo{booktitle}{{\em Deep Learning and Security
  Workshop (DLS)}}.
\newblock


\bibitem[\protect\citeauthoryear{Carlini and Wagner}{Carlini and
  Wagner}{2017b}]%
        {carlini2017towards}
\bibfield{author}{\bibinfo{person}{Nicholas Carlini} {and}
  \bibinfo{person}{David~A. Wagner}.} \bibinfo{year}{2017}\natexlab{b}.
\newblock \showarticletitle{{Towards Evaluating the Robustness of Neural
  Networks}}. In \bibinfo{booktitle}{{\em 2017 {IEEE} Symposium on Security and
  Privacy, {SP} 2017, San Jose, CA, USA, May 22-26, 2017}}.
  \bibinfo{pages}{39--57}.
\newblock
\showDOI{%
\url{https://doi.org/10.1109/SP.2017.49}}


\bibitem[\protect\citeauthoryear{Checkoway, McCoy, Kantor, Anderson, Shacham,
  Savage, Koscher, Czeskis, Roesner, and Kohno}{Checkoway
  et~al\mbox{.}}{2011}]%
        {vehicle-sec11}
\bibfield{author}{\bibinfo{person}{Stephen Checkoway}, \bibinfo{person}{Damon
  McCoy}, \bibinfo{person}{Brian Kantor}, \bibinfo{person}{Danny Anderson},
  \bibinfo{person}{Hovav Shacham}, \bibinfo{person}{Stefan Savage},
  \bibinfo{person}{Karl Koscher}, \bibinfo{person}{Alexei Czeskis},
  \bibinfo{person}{Franziska Roesner}, {and} \bibinfo{person}{Tadayoshi
  Kohno}.} \bibinfo{year}{2011}\natexlab{}.
\newblock \showarticletitle{{Comprehensive Experimental Analyses of Automotive
  Attack Surfaces}}. In \bibinfo{booktitle}{{\em Proceedings of the 20th USENIX
  Conference on Security}} {\em (\bibinfo{series}{SEC'11})}.
\newblock


\bibitem[\protect\citeauthoryear{Chen, Yin, Feng, Mao, and Liu}{Chen
  et~al\mbox{.}}{2018}]%
        {cvattack-ndss18}
\bibfield{author}{\bibinfo{person}{Qi~Alfred Chen}, \bibinfo{person}{Yucheng
  Yin}, \bibinfo{person}{Yiheng Feng}, \bibinfo{person}{Z.~Morley Mao}, {and}
  \bibinfo{person}{Henry X.~Liu Liu}.} \bibinfo{year}{2018}\natexlab{}.
\newblock \showarticletitle{{Exposing Congestion Attack on Emerging Connected
  Vehicle based Traffic Signal Control}}. In \bibinfo{booktitle}{{\em
  Proceedings of the 25th Annual Network and Distributed System Security
  Symposium}} {\em (\bibinfo{series}{NDSS '18})}.
\newblock


\bibitem[\protect\citeauthoryear{Cheng, Yi, Zhang, Chen, and Hsieh}{Cheng
  et~al\mbox{.}}{2018}]%
        {cheng2018seq2sick}
\bibfield{author}{\bibinfo{person}{Minhao Cheng}, \bibinfo{person}{Jinfeng Yi},
  \bibinfo{person}{Huan Zhang}, \bibinfo{person}{Pin-Yu Chen}, {and}
  \bibinfo{person}{Cho-Jui Hsieh}.} \bibinfo{year}{2018}\natexlab{}.
\newblock \showarticletitle{Seq2Sick: Evaluating the Robustness of
  Sequence-to-Sequence Models with Adversarial Examples}.
\newblock \bibinfo{journal}{{\em arXiv preprint arXiv:1803.01128\/}}
  (\bibinfo{year}{2018}).
\newblock


\bibitem[\protect\citeauthoryear{Cho and Shin}{Cho and Shin}{2016}]%
        {fault-vehicle-ccs16}
\bibfield{author}{\bibinfo{person}{Kyong-Tak Cho} {and}
  \bibinfo{person}{Kang~G. Shin}.} \bibinfo{year}{2016}\natexlab{}.
\newblock \showarticletitle{Error Handling of In-vehicle Networks Makes Them
  Vulnerable}. In \bibinfo{booktitle}{{\em Proceedings of the 2016 ACM SIGSAC
  Conference on Computer and Communications Security}} {\em
  (\bibinfo{series}{CCS'16})}.
\newblock


\bibitem[\protect\citeauthoryear{Cisse, Adi, Neverova, and Keshet}{Cisse
  et~al\mbox{.}}{2017}]%
        {cisse2017houdini}
\bibfield{author}{\bibinfo{person}{Moustapha Cisse}, \bibinfo{person}{Yossi
  Adi}, \bibinfo{person}{Natalia Neverova}, {and} \bibinfo{person}{Joseph
  Keshet}.} \bibinfo{year}{2017}\natexlab{}.
\newblock \showarticletitle{Houdini: Fooling deep structured prediction
  models}.
\newblock \bibinfo{journal}{{\em arXiv preprint arXiv:1707.05373\/}}
  (\bibinfo{year}{2017}).
\newblock


\bibitem[\protect\citeauthoryear{Evtimov, Eykholt, Fernandes, Kohno, Li,
  Prakash, Rahmati, and Song}{Evtimov et~al\mbox{.}}{2017}]%
        {evtimov2017robust}
\bibfield{author}{\bibinfo{person}{Ivan Evtimov}, \bibinfo{person}{Kevin
  Eykholt}, \bibinfo{person}{Earlence Fernandes}, \bibinfo{person}{Tadayoshi
  Kohno}, \bibinfo{person}{Bo Li}, \bibinfo{person}{Atul Prakash},
  \bibinfo{person}{Amir Rahmati}, {and} \bibinfo{person}{Dawn Song}.}
  \bibinfo{year}{2017}\natexlab{}.
\newblock \showarticletitle{Robust physical-world attacks on deep learning
  models}.
\newblock \bibinfo{journal}{{\em arXiv preprint arXiv:1707.08945\/}}
  \bibinfo{volume}{1} (\bibinfo{year}{2017}).
\newblock


\bibitem[\protect\citeauthoryear{Eykholt, Evtimov, Fernandes, Li, Rahmati,
  Tramer, Prakash, Kohno, and Song}{Eykholt et~al\mbox{.}}{2018a}]%
        {eykholt2018physical}
\bibfield{author}{\bibinfo{person}{Kevin Eykholt}, \bibinfo{person}{Ivan
  Evtimov}, \bibinfo{person}{Earlence Fernandes}, \bibinfo{person}{Bo Li},
  \bibinfo{person}{Amir Rahmati}, \bibinfo{person}{Florian Tramer},
  \bibinfo{person}{Atul Prakash}, \bibinfo{person}{Tadayoshi Kohno}, {and}
  \bibinfo{person}{Dawn Song}.} \bibinfo{year}{2018}\natexlab{a}.
\newblock \showarticletitle{{Physical Adversarial Examples for Object
  Detectors}}. In \bibinfo{booktitle}{{\em USENIX Workshop on Offensive
  Technologies (WOOT)}}.
\newblock


\bibitem[\protect\citeauthoryear{Eykholt, Evtimov, Fernandes, Li, Rahmati,
  Xiao, Prakash, Kohno, and Song}{Eykholt et~al\mbox{.}}{2018b}]%
        {eykholt2018robust}
\bibfield{author}{\bibinfo{person}{Kevin Eykholt}, \bibinfo{person}{Ivan
  Evtimov}, \bibinfo{person}{Earlence Fernandes}, \bibinfo{person}{Bo Li},
  \bibinfo{person}{Amir Rahmati}, \bibinfo{person}{Chaowei Xiao},
  \bibinfo{person}{Atul Prakash}, \bibinfo{person}{Tadayoshi Kohno}, {and}
  \bibinfo{person}{Dawn Song}.} \bibinfo{year}{2018}\natexlab{b}.
\newblock \showarticletitle{{Robust Physical-World Attacks on Deep Learning
  Visual Classification}}. In \bibinfo{booktitle}{{\em IEEE Conference on
  Computer Vision and Pattern Recognition (CVPR)}}.
\newblock


\bibitem[\protect\citeauthoryear{Feng, Huang, Chen, Liu, and Mao}{Feng
  et~al\mbox{.}}{2018}]%
        {trb:2018:yiheng:signalsecurity}
\bibfield{author}{\bibinfo{person}{Yiheng Feng}, \bibinfo{person}{Shihong
  Huang}, \bibinfo{person}{Qi~Alfred Chen}, \bibinfo{person}{Henry~X. Liu},
  {and} \bibinfo{person}{Z.~Morley Mao}.} \bibinfo{year}{2018}\natexlab{}.
\newblock \showarticletitle{{Vulnerability of Traffic Control System Under
  Cyber-Attacks Using Falsified Data}}. In \bibinfo{booktitle}{{\em
  Transportation Research Board 2018 Annual Meeting (TRB)}}.
\newblock


\bibitem[\protect\citeauthoryear{Goodfellow, Shlens, and Szegedy}{Goodfellow
  et~al\mbox{.}}{2014}]%
        {goodfellow2014explaining}
\bibfield{author}{\bibinfo{person}{Ian~J. Goodfellow},
  \bibinfo{person}{Jonathon Shlens}, {and} \bibinfo{person}{Christian
  Szegedy}.} \bibinfo{year}{2014}\natexlab{}.
\newblock \showarticletitle{Explaining and harnessing adversarial examples}.
\newblock \bibinfo{journal}{{\em arXiv preprint arXiv:1412.6572\/}}
  (\bibinfo{year}{2014}).
\newblock


\bibitem[\protect\citeauthoryear{Inc.}{Inc.}{2018}]%
        {vpl}
\bibfield{author}{\bibinfo{person}{Velodyne~LiDAR Inc.}}
  \bibinfo{year}{2018}\natexlab{}.
\newblock \bibinfo{title}{{VLP-16 User Manual}}.
\newblock   (\bibinfo{year}{2018}).
\newblock


\bibitem[\protect\citeauthoryear{Ivanov, Pajic, and Lee}{Ivanov
  et~al\mbox{.}}{2014}]%
        {ivanov2014attack}
\bibfield{author}{\bibinfo{person}{Radoslav Ivanov}, \bibinfo{person}{Miroslav
  Pajic}, {and} \bibinfo{person}{Insup Lee}.} \bibinfo{year}{2014}\natexlab{}.
\newblock \showarticletitle{Attack-resilient sensor fusion}. In
  \bibinfo{booktitle}{{\em Proceedings of the conference on Design, Automation
  \& Test in Europe}}.
\newblock


\bibitem[\protect\citeauthoryear{Jaderberg, Simonyan, Zisserman,
  et~al\mbox{.}}{Jaderberg et~al\mbox{.}}{2015}]%
        {jaderberg2015spatial}
\bibfield{author}{\bibinfo{person}{Max Jaderberg}, \bibinfo{person}{Karen
  Simonyan}, \bibinfo{person}{Andrew Zisserman}, {et~al\mbox{.}}}
  \bibinfo{year}{2015}\natexlab{}.
\newblock \showarticletitle{Spatial transformer networks}. In
  \bibinfo{booktitle}{{\em Advances in neural information processing systems}}.
  \bibinfo{pages}{2017--2025}.
\newblock


\bibitem[\protect\citeauthoryear{Kingma and Ba}{Kingma and Ba}{2014}]%
        {kingma2014adam}
\bibfield{author}{\bibinfo{person}{Diederik~P Kingma} {and}
  \bibinfo{person}{Jimmy Ba}.} \bibinfo{year}{2014}\natexlab{}.
\newblock \showarticletitle{Adam: A method for stochastic optimization}.
\newblock \bibinfo{journal}{{\em arXiv preprint arXiv:1412.6980\/}}
  (\bibinfo{year}{2014}).
\newblock


\bibitem[\protect\citeauthoryear{Koscher, Czeskis, Roesner, Patel, Kohno,
  Checkoway, McCoy, Kantor, Anderson, Shacham, and Savage}{Koscher
  et~al\mbox{.}}{2010}]%
        {vehicle-oakland10}
\bibfield{author}{\bibinfo{person}{Karl Koscher}, \bibinfo{person}{Alexei
  Czeskis}, \bibinfo{person}{Franziska Roesner}, \bibinfo{person}{Shwetak
  Patel}, \bibinfo{person}{Tadayoshi Kohno}, \bibinfo{person}{Stephen
  Checkoway}, \bibinfo{person}{Damon McCoy}, \bibinfo{person}{Brian Kantor},
  \bibinfo{person}{Danny Anderson}, \bibinfo{person}{Hovav Shacham}, {and}
  \bibinfo{person}{Stefan Savage}.} \bibinfo{year}{2010}\natexlab{}.
\newblock \showarticletitle{{Experimental Security Analysis of a Modern
  Automobile}}. In \bibinfo{booktitle}{{\em Proceedings of the 2010 IEEE
  Symposium on Security and Privacy}} {\em (\bibinfo{series}{SP'10})}.
\newblock


\bibitem[\protect\citeauthoryear{Liu, Ma, Aafer, Lee, Zhai, Wang, and
  Zhang}{Liu et~al\mbox{.}}{2018}]%
        {liu2017trojaning}
\bibfield{author}{\bibinfo{person}{Yingqi Liu}, \bibinfo{person}{Shiqing Ma},
  \bibinfo{person}{Yousra Aafer}, \bibinfo{person}{Wen-Chuan Lee},
  \bibinfo{person}{Juan Zhai}, \bibinfo{person}{Weihang Wang}, {and}
  \bibinfo{person}{Xiangyu Zhang}.} \bibinfo{year}{2018}\natexlab{}.
\newblock \showarticletitle{{Trojaning Attack on Neural Networks}}. In
  \bibinfo{booktitle}{{\em Annual Network and Distributed System Security
  Symposium (NDSS)}}.
\newblock


\bibitem[\protect\citeauthoryear{Ma, Li, Wang, Erfani, Wijewickrema, Houle,
  Schoenebeck, Song, and Bailey}{Ma et~al\mbox{.}}{2018}]%
        {ma2018characterizing}
\bibfield{author}{\bibinfo{person}{Xingjun Ma}, \bibinfo{person}{Bo Li},
  \bibinfo{person}{Yisen Wang}, \bibinfo{person}{Sarah~M Erfani},
  \bibinfo{person}{Sudanthi Wijewickrema}, \bibinfo{person}{Michael~E Houle},
  \bibinfo{person}{Grant Schoenebeck}, \bibinfo{person}{Dawn Song}, {and}
  \bibinfo{person}{James Bailey}.} \bibinfo{year}{2018}\natexlab{}.
\newblock \showarticletitle{Characterizing Adversarial Subspaces Using Local
  Intrinsic Dimensionality}.
\newblock \bibinfo{journal}{{\em arXiv preprint arXiv:1801.02613\/}}
  (\bibinfo{year}{2018}).
\newblock


\bibitem[\protect\citeauthoryear{Madry, Makelov, Schmidt, Tsipras, and
  Vladu}{Madry et~al\mbox{.}}{2017}]%
        {madry2017towards}
\bibfield{author}{\bibinfo{person}{Aleksander Madry},
  \bibinfo{person}{Aleksandar Makelov}, \bibinfo{person}{Ludwig Schmidt},
  \bibinfo{person}{Dimitris Tsipras}, {and} \bibinfo{person}{Adrian Vladu}.}
  \bibinfo{year}{2017}\natexlab{}.
\newblock \showarticletitle{Towards deep learning models resistant to
  adversarial attacks}.
\newblock \bibinfo{journal}{{\em arXiv preprint arXiv:1706.06083\/}}
  (\bibinfo{year}{2017}).
\newblock


\bibitem[\protect\citeauthoryear{Mazloom, Rezaeirad, Hunter, and McCoy}{Mazloom
  et~al\mbox{.}}{2016}]%
        {mazloom2016security}
\bibfield{author}{\bibinfo{person}{Sahar Mazloom}, \bibinfo{person}{Mohammad
  Rezaeirad}, \bibinfo{person}{Aaron Hunter}, {and} \bibinfo{person}{Damon
  McCoy}.} \bibinfo{year}{2016}\natexlab{}.
\newblock \showarticletitle{{A Security Analysis of an In-Vehicle Infotainment
  and App Platform}}. In \bibinfo{booktitle}{{\em Usenix Workshop on Offensive
  Technologies (WOOT)}}.
\newblock


\bibitem[\protect\citeauthoryear{Papernot, McDaniel, Goodfellow, Jha, Celik,
  and Swami}{Papernot et~al\mbox{.}}{2017}]%
        {papernot2017practical}
\bibfield{author}{\bibinfo{person}{Nicolas Papernot}, \bibinfo{person}{Patrick
  McDaniel}, \bibinfo{person}{Ian Goodfellow}, \bibinfo{person}{Somesh Jha},
  \bibinfo{person}{Z~Berkay Celik}, {and} \bibinfo{person}{Ananthram Swami}.}
  \bibinfo{year}{2017}\natexlab{}.
\newblock \showarticletitle{{Practical Black-Box Attacks Against Machine
  Learning}}. In \bibinfo{booktitle}{{\em ACM on Asia Conference on Computer
  and Communications Security}}.
\newblock


\bibitem[\protect\citeauthoryear{Petit, Stottelaar, Feiri, and Kargl}{Petit
  et~al\mbox{.}}{2015}]%
        {sensor-blackhat15}
\bibfield{author}{\bibinfo{person}{Jonathan Petit}, \bibinfo{person}{Bas
  Stottelaar}, \bibinfo{person}{Michael Feiri}, {and} \bibinfo{person}{Frank
  Kargl}.} \bibinfo{year}{2015}\natexlab{}.
\newblock \showarticletitle{{Remote Attacks on Automated Vehicles Sensors:
  Experiments on Camera and LiDAR}}. In \bibinfo{booktitle}{{\em Black Hat
  Europe}}.
\newblock


\bibitem[\protect\citeauthoryear{Rouf, Miller, Mustafa, Taylor, Oh, Xu,
  Gruteser, Trappe, and Seskar}{Rouf et~al\mbox{.}}{2010}]%
        {RoufTPMSAttack}
\bibfield{author}{\bibinfo{person}{Ishtiaq Rouf}, \bibinfo{person}{Rob Miller},
  \bibinfo{person}{Hossen Mustafa}, \bibinfo{person}{Travis Taylor},
  \bibinfo{person}{Sangho Oh}, \bibinfo{person}{Wenyuan Xu},
  \bibinfo{person}{Marco Gruteser}, \bibinfo{person}{Wade Trappe}, {and}
  \bibinfo{person}{Ivan Seskar}.} \bibinfo{year}{2010}\natexlab{}.
\newblock \showarticletitle{Security and Privacy Vulnerabilities of In-car
  Wireless Networks: A Tire Pressure Monitoring System Case Study}. In
  \bibinfo{booktitle}{{\em Proceedings of the 19th USENIX Conference on
  Security}} {\em (\bibinfo{series}{USENIX Security'10})}.
  \bibinfo{publisher}{USENIX Association}, \bibinfo{address}{Berkeley, CA,
  USA}, \bibinfo{pages}{21--21}.
\newblock
\showISBNx{888-7-6666-5555-4}
\showURL{%
\url{http://dl.acm.org/citation.cfm?id=1929820.1929848}}


\bibitem[\protect\citeauthoryear{Shin, Kim, Kwon, and Kim}{Shin
  et~al\mbox{.}}{2017}]%
        {Shin2017IllusionAD}
\bibfield{author}{\bibinfo{person}{Hocheol Shin}, \bibinfo{person}{Dohyun Kim},
  \bibinfo{person}{Yujin Kwon}, {and} \bibinfo{person}{Yongdae Kim}.}
  \bibinfo{year}{2017}\natexlab{}.
\newblock \showarticletitle{{Illusion and Dazzle: Adversarial Optical Channel
  Exploits Against Lidars for Automotive Applications}}. In
  \bibinfo{booktitle}{{\em International Conference on Cryptographic Hardware
  and Embedded Systems (CHES)}}.
\newblock


\bibitem[\protect\citeauthoryear{Shoukry, Martin, Tabuada, and
  Srivastava}{Shoukry et~al\mbox{.}}{2013}]%
        {ShoukryABSSpoofing}
\bibfield{author}{\bibinfo{person}{Yasser Shoukry}, \bibinfo{person}{Paul
  Martin}, \bibinfo{person}{Paulo Tabuada}, {and} \bibinfo{person}{Mani
  Srivastava}.} \bibinfo{year}{2013}\natexlab{}.
\newblock \showarticletitle{Non-invasive Spoofing Attacks for Anti-lock Braking
  Systems}. In \bibinfo{booktitle}{{\em Cryptographic Hardware and Embedded
  Systems - CHES 2013}}, \bibfield{editor}{\bibinfo{person}{Guido Bertoni}
  {and} \bibinfo{person}{Jean-S{\'e}bastien Coron}} (Eds.).
  \bibinfo{publisher}{Springer Berlin Heidelberg}, \bibinfo{address}{Berlin,
  Heidelberg}, \bibinfo{pages}{55--72}.
\newblock
\showISBNx{978-3-642-40349-1}


\bibitem[\protect\citeauthoryear{Shoukry, Martin, Yona, Diggavi, and
  Srivastava}{Shoukry et~al\mbox{.}}{2015}]%
        {Shoukry2015PyCRAPC}
\bibfield{author}{\bibinfo{person}{Yasser Shoukry}, \bibinfo{person}{Paul
  Martin}, \bibinfo{person}{Yair Yona}, \bibinfo{person}{Suhas~N. Diggavi},
  {and} \bibinfo{person}{Mani~B. Srivastava}.} \bibinfo{year}{2015}\natexlab{}.
\newblock \showarticletitle{PyCRA: Physical Challenge-Response Authentication
  For Active Sensors Under Spoofing Attacks}. In \bibinfo{booktitle}{{\em ACM
  Conference on Computer and Communications Security}}.
\newblock


\bibitem[\protect\citeauthoryear{Tram{\`e}r, Kurakin, Papernot, Boneh, and
  McDaniel}{Tram{\`e}r et~al\mbox{.}}{2017}]%
        {tramer2017ensemble}
\bibfield{author}{\bibinfo{person}{Florian Tram{\`e}r}, \bibinfo{person}{Alexey
  Kurakin}, \bibinfo{person}{Nicolas Papernot}, \bibinfo{person}{Dan Boneh},
  {and} \bibinfo{person}{Patrick McDaniel}.} \bibinfo{year}{2017}\natexlab{}.
\newblock \showarticletitle{Ensemble adversarial training: Attacks and
  defenses}.
\newblock \bibinfo{journal}{{\em arXiv preprint arXiv:1705.07204\/}}
  (\bibinfo{year}{2017}).
\newblock


\bibitem[\protect\citeauthoryear{Wong, Huang, Feng, Chen, Mao, and Liu}{Wong
  et~al\mbox{.}}{2019}]%
        {wong2019trajectory}
\bibfield{author}{\bibinfo{person}{Wai Wong}, \bibinfo{person}{Shihong Huang},
  \bibinfo{person}{Yiheng Feng}, \bibinfo{person}{Qi~Alfred Chen},
  \bibinfo{person}{Z~Morley Mao}, {and} \bibinfo{person}{Henry~X Liu}.}
  \bibinfo{year}{2019}\natexlab{}.
\newblock \showarticletitle{{Trajectory-Based Hierarchical Defense Model to
  Detect Cyber-Attacks on Transportation Infrastructure}}. In
  \bibinfo{booktitle}{{\em Transportation Research Board 2019 Annual Meeting
  (TRB)}}.
\newblock


\bibitem[\protect\citeauthoryear{Xiang, Qi, and Li}{Xiang
  et~al\mbox{.}}{2018}]%
        {xiang2018generating}
\bibfield{author}{\bibinfo{person}{Chong Xiang}, \bibinfo{person}{Charles~R
  Qi}, {and} \bibinfo{person}{Bo Li}.} \bibinfo{year}{2018}\natexlab{}.
\newblock \showarticletitle{Generating 3D Adversarial Point Clouds}.
\newblock \bibinfo{journal}{{\em arXiv preprint arXiv:1809.07016\/}}
  (\bibinfo{year}{2018}).
\newblock


\bibitem[\protect\citeauthoryear{Xiao, Deng, Li, Yu, Song, et~al\mbox{.}}{Xiao
  et~al\mbox{.}}{2018a}]%
        {xiao2018characterizing}
\bibfield{author}{\bibinfo{person}{Chaowei Xiao}, \bibinfo{person}{Ruizhi
  Deng}, \bibinfo{person}{Bo Li}, \bibinfo{person}{Fisher Yu},
  \bibinfo{person}{Dawn Song}, {et~al\mbox{.}}}
  \bibinfo{year}{2018}\natexlab{a}.
\newblock \showarticletitle{Characterizing Adversarial Examples Based on
  Spatial Consistency Information for Semantic Segmentation}. In
  \bibinfo{booktitle}{{\em Proceedings of the (ECCV)}}.
  \bibinfo{pages}{217--234}.
\newblock


\bibitem[\protect\citeauthoryear{Xiao, Li, Zhu, He, Liu, and Song}{Xiao
  et~al\mbox{.}}{2018b}]%
        {xiao2018generating}
\bibfield{author}{\bibinfo{person}{Chaowei Xiao}, \bibinfo{person}{Bo Li},
  \bibinfo{person}{Jun-Yan Zhu}, \bibinfo{person}{Warren He},
  \bibinfo{person}{Mingyan Liu}, {and} \bibinfo{person}{Dawn Song}.}
  \bibinfo{year}{2018}\natexlab{b}.
\newblock \showarticletitle{Generating adversarial examples with adversarial
  networks}.
\newblock \bibinfo{journal}{{\em arXiv preprint arXiv:1801.02610\/}}
  (\bibinfo{year}{2018}).
\newblock


\bibitem[\protect\citeauthoryear{Xiao, Zhu, Li, He, Liu, and Song}{Xiao
  et~al\mbox{.}}{2018c}]%
        {xiao2018spatially}
\bibfield{author}{\bibinfo{person}{Chaowei Xiao}, \bibinfo{person}{Jun-Yan
  Zhu}, \bibinfo{person}{Bo Li}, \bibinfo{person}{Warren He},
  \bibinfo{person}{Mingyan Liu}, {and} \bibinfo{person}{Dawn Song}.}
  \bibinfo{year}{2018}\natexlab{c}.
\newblock \showarticletitle{Spatially transformed adversarial examples}.
\newblock \bibinfo{journal}{{\em arXiv preprint arXiv:1801.02612\/}}
  (\bibinfo{year}{2018}).
\newblock


\bibitem[\protect\citeauthoryear{Xie, Wang, Zhang, Zhou, Xie, and Yuille}{Xie
  et~al\mbox{.}}{2017}]%
        {xie2017adversarial}
\bibfield{author}{\bibinfo{person}{Cihang Xie}, \bibinfo{person}{Jianyu Wang},
  \bibinfo{person}{Zhishuai Zhang}, \bibinfo{person}{Yuyin Zhou},
  \bibinfo{person}{Lingxi Xie}, {and} \bibinfo{person}{Alan Yuille}.}
  \bibinfo{year}{2017}\natexlab{}.
\newblock \showarticletitle{{Adversarial Examples for Semantic Segmentation and
  Object Detection}}. In \bibinfo{booktitle}{{\em IEEE International Conference
  on Computer Vision (ICCV)}}.
\newblock


\bibitem[\protect\citeauthoryear{Xu, Chen, Liu, Rohrbach, Darell, and Song}{Xu
  et~al\mbox{.}}{2017}]%
        {xu2017can}
\bibfield{author}{\bibinfo{person}{Xiaojun Xu}, \bibinfo{person}{Xinyun Chen},
  \bibinfo{person}{Chang Liu}, \bibinfo{person}{Anna Rohrbach},
  \bibinfo{person}{Trevor Darell}, {and} \bibinfo{person}{Dawn Song}.}
  \bibinfo{year}{2017}\natexlab{}.
\newblock \showarticletitle{Can you fool AI with adversarial examples on a
  visual Turing test?}
\newblock \bibinfo{journal}{{\em arXiv preprint arXiv:1709.08693\/}}
  (\bibinfo{year}{2017}).
\newblock


\bibitem[\protect\citeauthoryear{Yan}{Yan}{2016}]%
        {Yan2016CanYT}
\bibfield{author}{\bibinfo{person}{Chen Yan}.} \bibinfo{year}{2016}\natexlab{}.
\newblock \showarticletitle{Can You Trust Autonomous Vehicles : Contactless
  Attacks against Sensors of Self-driving Vehicle}.
\newblock


\bibitem[\protect\citeauthoryear{Yang, Wang, Jiang, Song, Luo, Guan, Li, and
  Shi}{Yang et~al\mbox{.}}{2018}]%
        {yang2018sensor}
\bibfield{author}{\bibinfo{person}{Kang Yang}, \bibinfo{person}{Rui Wang},
  \bibinfo{person}{Yu Jiang}, \bibinfo{person}{Houbing Song},
  \bibinfo{person}{Chenxia Luo}, \bibinfo{person}{Yong Guan},
  \bibinfo{person}{Xiaojuan Li}, {and} \bibinfo{person}{Zhiping Shi}.}
  \bibinfo{year}{2018}\natexlab{}.
\newblock \showarticletitle{Sensor attack detection using history based
  pairwise inconsistency}.
\newblock \bibinfo{journal}{{\em Future Generation Computer Systems\/}}
  (\bibinfo{year}{2018}).
\newblock


\bibitem[\protect\citeauthoryear{Yuan, Chen, Zhao, Long, Liu, Chen, Zhang,
  Huang, Wang, and Gunter}{Yuan et~al\mbox{.}}{2018}]%
        {yuan2018commandersong}
\bibfield{author}{\bibinfo{person}{Xuejing Yuan}, \bibinfo{person}{Yuxuan
  Chen}, \bibinfo{person}{Yue Zhao}, \bibinfo{person}{Yunhui Long},
  \bibinfo{person}{Xiaokang Liu}, \bibinfo{person}{Kai Chen},
  \bibinfo{person}{Shengzhi Zhang}, \bibinfo{person}{Heqing Huang},
  \bibinfo{person}{Xiaofeng Wang}, {and} \bibinfo{person}{Carl~A Gunter}.}
  \bibinfo{year}{2018}\natexlab{}.
\newblock \showarticletitle{{CommanderSong: A Systematic Approach for Practical
  Adversarial Voice Recognition}}. In \bibinfo{booktitle}{{\em USENIX Security
  Symposium}}.
\newblock


\end{thebibliography}
\appendix
\section*{Appendix}
\section{Algorithm Details and Experiment Settings}
\label{appendix:algorithm}
Algorithm~\ref{alg1} shows the detailed algorithm to generate adversarial examples. In our experiment, we selected Adam~\cite{kingma2014adam} as our optimizer $opt$ with learning rate $1e-4$. $opt(l_{\adv}; \theta, \tau_x, s_h)$ means updating the parameters  $\theta, \tau_x, s_h)$ with respect to Loss function $l_{\adv}$. We selected TensorFlow~\cite{abadi2016tensorflow} as our backbone. $L_{t}$ is set as $12.5$ while $L_{\theta}$ is set as the angle that generates $2$-meter distance from the target position.


\begin{algorithm}[bh!]
\SetAlgoLined
\SetKwInput{Input}{input}
\SetKwInput{Output}{output}
\SetKwInput{Init}{Initialization}
\SetKwInput{Blank}{}
\SetKwInput{Ret}{Return}

\tabcolsep=0pt
\begin{tabular}{@{}ll}
    \Input{}&Target model: $\boldsymbol{M}$\;\\
&  \pointcloud{} $\boldsymbol{X}$  \;\\
&3D \advsensor{} $\boldsymbol{T}$\;\\
& Optimizer $\boldsymbol{opt}$\;\\
& Max iteration $\boldsymbol{N}$\;\\
    \Output{}&3D adversarial \pointcloud{} $\boldsymbol{X'}$\;\\
\end{tabular}
\BlankLine
 $\V{Initialization}$: $\theta \gets 0$, $\tau_x \gets 0$, $s_h \gets 1$, $l_{min} = +\mathrm{inf}$, $x = \Phi(X), t = \Phi(T)$\;
 \tcc{Initiate parameters by sampling around the transformation parameters $Target_{\theta}$, $Target_{\tau_x}$ that transforms t to the target position $(px,py)$ of the attack}
 \For{$i\tau_x \leftarrow -L_t$ \KwTo $L_t$}{
 \For{$i\theta \leftarrow -L_{\theta}$ \KwTo $L_{\theta}$}{
 \tcc*[h]{Initialize parameter .}\;
 $\theta \gets Target_{\theta} + i\theta$, $\tau_x \gets Target_{\tau_x} + \tau_xi$\;
 \For{$iter \leftarrow 1$ \KwTo $N$}{
 \tcc*[h]{Calculate adversarial loss }\;
 $l_{\adv} \gets \V{Equation}$~\ref{eq:adv}.\;
 \tcc{Update the parameters $\theta,\tau_x, s_h$ based on optimizer $opt$ and loss $l_{\adv}$ }\;
 $\theta,\tau_x,s_h \gets opt(l_{\adv};\theta,\tau_x,s_h) $

  \uIf{ $l_{min} < l_{adv}$}{
    $\theta^{final},\tau_x^{final},s_h^{final} \gets \theta,\tau_x,s_h $ 
  }
}
}
}
$T'\gets G_T(\theta^{final},t_x^{final},s_h^{final} ; T)$\;
$X' \gets X+ T'$\;
\Ret{$\advdelta$}
 \caption{Generating adversarial examples by leveraging global spatial transformation }
 \label{alg1}
\end{algorithm}

\begin{figure}[bh!]
  \centering
    \includegraphics[width=0.4\textwidth]{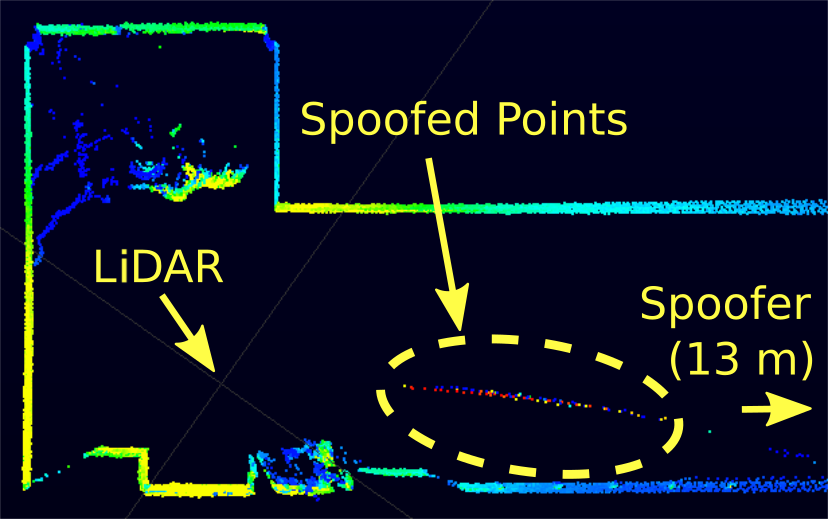}
    \caption{Collected traces from the reproduced sensor attack. The points in the yellow circle are spoofed by the sensor attack.}
    \label{fig:LiDAR_sensor_attack}
\end{figure}

\end{document}